\documentclass[onecolumn,aps,nofootinbib,a4paper]{revtex4} 

\usepackage[english]{babel}
\usepackage[utf8]{inputenc}

\usepackage{amsmath,amssymb}
\usepackage{graphicx}

\newcommand{\ie}{i.\,e.\ }
\newcommand{\cf}{cf.\ }
\newcommand{\etc}{etc.\ }
\newcommand{\etal}{et\,al.\ }
\newcommand{\dd}{\text{d}}
\newcommand{\re}{\operatorname{Re}}
\newcommand{\im}{\operatorname{Im}}

\newcommand{\braket}[2]{\langle{#1}|{#2}\rangle}
\newcommand{\bkew}[3]{\langle{#1}|{#2}|{#3}\rangle}
\newcommand{\bkewreduced}[3] {\langle{#1}||{#2}||{#3}\rangle}

\newcommand{\ket}[1]{\left|{#1}\right\rangle}

\newcommand{\ketbra}[2]{\left|{#1}\middle\rangle\middle\langle{#2}\right|}

\newcommand{\ew}[1]{\langle{#1}\rangle}
\newcommand{\ewbig}[1]{\big\langle{#1}\big\rangle}

\newcommand{\vct}[1]{\mathbf{#1}}

\newcommand{\abs}[1]{\lvert#1\rvert}

\newcommand{\tder}[2]{\tfrac{d^{#2}}{d {#1}^{#2}}}
\newcommand{\pder}[2]{\frac{\partial^{#2}}{\partial {#1}^{#2}}}
\newcommand{\tpder}[2]{\tfrac{\partial^{#2}}{\partial {#1}^{#2}}}

\newcommand{\bkappa}{\boldsymbol{\kappa}}
\newcommand{\bepsilon}{\boldsymbol{\epsilon}}
\newcommand{\be}{\mathbf{e}}

\newcommand{\bk}{\mathbf{k}}
\newcommand{\br}{\mathbf{r}}
\newcommand{\bp}{\mathbf{p}}

\newcommand{\cc}{\mathrm{c.\,c.}}
\newcommand{\Hc}{\mathrm{H.\,c.}}

\begin{document}
%
%
\title{The R\"{o}ntgen interaction and forces on dipoles in time-modulated optical fields}
\author{Matthias Sonnleitner}
\affiliation{School of Physics and Astronomy, University of Glasgow, Glasgow G12~8QQ, United Kingdom}
\author{Stephen M. Barnett}
\affiliation{School of Physics and Astronomy, University of Glasgow, Glasgow G12~8QQ, United Kingdom}
\begin{abstract}
	 The R\"{o}ntgen term is an often neglected contribution to the interaction between an atom and an electromagnetic field in the electric dipole approximation. In this work we discuss how this interaction term leads to a difference between the kinetic and canonical momentum of an atom which, in turn, leads to surprising radiation forces acting on the atom. We use a number of examples to explore the main features of this interaction, namely forces acting against the expected dipole force or accelerations perpendicular to the beam propagation axis.
\end{abstract}

%
\maketitle
%
%
\section{Introduction}
The mechanical interaction between light and atoms, molecules, nano- or micro-particles has become a key ingredient to many experiments and technologies in (quantum-) optics, chemistry and biology. These forces are not only used to trap, cool and move particles~\cite{cohen1998nobel,phillips1998nobel,ashkin2006optical}, they can also be controlled to unprecedented level to test other phenomena of interest, such as, for instance, gravitational waves~\cite{abbott2016observation} and even proposed models for dark-energy~\cite{burrage2015probing,hamilton2015atom}.

With experiments pushing the frontier of measurement and control it is time to revisit the theory of light-matter interaction and reconsider terms previously deemed negligible. One of these often neglected terms is the so-called R\"{o}ntgen term which extends the interaction between an atom and electromagnetic fields in the electric dipole approximation from $H_\text{AL}=- \vct{d}\cdot\vct{E}$ to 
\begin{equation}\label{eq:_Hatomlaser_into}
	H_\text{AL} = - \vct{d}\cdot\vct{E} - \tfrac{1}{2 M} \big( \vct{P} \cdot ( \vct{B}\times\vct{d}) + ( \vct{B}\times\vct{d})\cdot \vct{P}\big)	\,.
\end{equation}
The R\"{o}ntgen term is named after Wilhelm Conrad R\"{o}ntgen, who observed a magnetic field arising from a dielectric moving through an electric field~\cite{rontgen1888ueber}. It accounts for the fact that a moving electric dipole appears to carry a magnetic dipole moment which interacts with a magnetic field~\cite{jackson1975classical,wilkens1994quantum,hnizdo2012magnetic}. 

This term appears naturally in a rigorous quantum-electrodynamical derivation of the interaction Hamiltonian between atoms and electromagnetic fields~\cite{baxter1993canonical,lembessis1993theory,kozlovskii2008doppler,power1964introductory}. It has received some attention in connection with the spontaneous decay of moving atoms~\cite{wilkens1994significance,cresser2003rate,sonnleitner2017will}, the so-called Abraham-Minkowski-controversy~\cite{hinds2009momentum,barnett2010enigma,barnett2010resolution} and as a source of Aharonov-Bohm-type phase shifts for moving dipoles~\cite{wilkens1994quantum,leonhardt1999quantized,horsley2005rontgen}.

One of the most intriguing features of the Röntgen term is that it induces a momentum-dependent interaction~$H_\text{AL}$. As outlined in Sect.~\ref{sec:_intro_class_discussion}, this leads to a difference between the canonical momentum~$\vct{P}$ and the mechanical momentum~$M \dot{\vct{R}}$. Especially if properties of the field, such as the amplitude or the phase, are modulated in time, these different momenta can give rise to force terms which would not appear in the usual $\vct{d}\cdot\vct{E}$--interaction. Although this feature has been noticed before, for instance in refs.~\cite{lembessis1993theory} or~\cite{hinds2009momentum}, a systematic study of these effects is desirable.

The purpose of this work is to study radiation force terms arising due to the presence of the Röntgen term in the context of time-modulated laser fields. We show that these forces can show counter-intuitive behaviour: usually high-field-seeking atoms are pushed away from the maximum of intensity of a travelling laser pulse and special configurations allow for forces perpendicular to the propagation axis of a transversally homogeneous laser beam.

The outline of this work is as follows: we first use a classical argument to introduce the Röntgen-term and the corresponding forces in Sect.~\ref{sec:_intro_class_discussion}. In Sect.~\ref{sec:_2lvlexamples} we go on to discuss several fundamental characteristics using the example of a two-level atom interacting with an amplitude-modulated laser beam. Finally we shall explore options from more evolved setups in Sect.~\ref{sec:_4lvl_examples}, followed by a discussion and conclusions in sections~\ref{sec:_Discussion_HowSmall} and~\ref{sec:_Conclusion}.

As we try to explore and discuss generic properties of these forces we stick to a semi-classical treatment for an atom moving in a classical laser field. The expectation values of the atom's position and momentum are treated classically and momentum-diffusion due to spontaneous emission is ignored. But we will also see that the forces described here are, in general, sufficiently small such that quantum effects should be considered to fully describe a specific experiment. We believe, however, that a semi-classical approach is the more intuitive way of providing a broad overview.

\section{Short classical motivation of the Röntgen force}\label{sec:_intro_class_discussion}

Let us start with a short classical motivation of the Röntgen interaction between a classical electric dipole and a magnetic field and the associated forces, which shall be explored in more detail the rest of this work.

The interaction between a particle of mass $M$ with an electric and magnetic dipole moment with electric or magnetic fields is described by the Lagrangian
\begin{equation}
	L = \tfrac{1}{2} M \vct{v}^2 + \vct{d} \cdot \vct{E} + \vct{m} \cdot \vct{B} \,.
\end{equation}
The dipole moments $\vct{d}$ and $\vct{m}$ here appear as measured in the lab frame where the atom moves with a velocity~$\vct{v}=\dot{\vct{r}}$. To first order in $\abs{\vct{v}}/c$ they are connected to the corresponding quantities in the atom's rest frame by $\vct{d} = \vct{d}' + \tfrac{1}{c^2} \vct{v}\times\vct{m}'$ and $\vct{m} = \vct{m}' - \vct{v}\times\vct{d}'$~\cite{hnizdo2012magnetic}.

A purely dielectric particle with $\vct{m}'=0$, \ie $\vct{d} = \vct{d}'$, will therefore interact with the magnetic field in the lab frame as $\vct{m}=- \vct{v}\times\vct{d}$ such that
\begin{equation}
	L = \tfrac{1}{2} M \vct{v}^2 + \vct{d} \cdot \vct{E} + \vct{v} \cdot \big(\vct{B} \times \vct{d}\big)	\,.
\end{equation}
The canonical momentum $\vct{p} = \tpder{\dot{\vct{r}}}{} L = \tpder{\vct{v}}{} L$ is thus different from the kinetic momentum $M \vct{v}$ as
\begin{equation}\label{eq:_kinetic_canonic_momentum}
	\vct{p} = M \vct{v} + \vct{B}\times\vct{d}	\,.
\end{equation}
The Hamiltonian given in Eq.~\eqref{eq:_Hatomlaser_into} can then be obtained using a Legendre transformation $H = \vct{p} \cdot \vct{v} - L$ after dropping terms~$\sim (\vct{B}\times\vct{d})^2$ and using that $\vct{p}$ and $\vct{B}$ commute in classical mechanics.

The equations of motion now show that $\dot{\bp} = -\nabla\big[ \vct{d} \cdot \vct{E} + \vct{v} \cdot \big(\vct{B} \times \vct{d}\big)\big]$, hence the Röntgen-term gives a velocity-dependent correction to the change in canonical momentum. But another intriguing feature arises from the difference between the change in canonical momentum and the kinetic force actually acting on the particle,
\begin{equation}\label{eq:_kinetic_canonic_force}
	M \ddot{\br} = \dot{\bp} - \tder{t}{} \big[\vct{B}\times\vct{d}\big]	\,.
\end{equation}
Hence, although the Röntgen-term has been introduced as a velocity-dependent interaction, it can give rise to a force acting on particles \emph{at rest}, if $\vct{B}\times\vct{d}$ changes in time. A corresponding expression can also be derived by simply adding up the Lorentz force acting on the two opposite charges constituting the dipole~\cite{stenholm1986semiclassical}.

A similar mechanism with magnetic dipoles, that is with $\vct{m}'\neq 0$ but $\vct{d}'=0$, results in the Aharonov-Casher effect~\cite{aharonov1984topological,cimmino1989observation,sangster1993measurement,konig2006direct}. One can connect these special cases with the more general expression $\vct{p} = M \vct{v} + \vct{B}\times\vct{d} - \vct{E}\times \vct{m}$ using the duality-- or Heavyside--Larmor--transformation,
\begin{subequations}
\begin{align}
	\vct{E} &\rightarrow \vct{E}' \cos\theta + c \mu_0 \vct{H} \sin\theta \,, \\
	\vct{H} &\rightarrow \vct{H}' \cos\theta - c \varepsilon_0 \vct{E}' \sin\theta \,, \\
	\vct{D} &\rightarrow \vct{D}' \cos\theta + c \varepsilon_0 \vct{B}' \sin\theta \,, \\
	\vct{B} &\rightarrow \vct{B}' \cos\theta -c \mu_0 \vct{D}' \sin\theta \,, \\
	\vct{d} &\rightarrow \vct{d}' \cos\theta + c^{-1} \vct{m}' \sin\theta \,, \\
	\vct{m} &\rightarrow \vct{m}' \cos\theta - c \vct{d}' \sin\theta \,,
\end{align}
\end{subequations}
for a real angle $\theta$~\cite{jackson1975classical,cameron2012electric}. The discrepancy between the canonical and kinetic momenta for matter interacting with electromagnetic fields lies at the core of controversies regarding the momentum of light or so-called hidden momenta~\cite{barnett2010enigma,saldanha2016hidden}.

In this work we discuss forces arising from classical laser fields interacting with an atom in the electric-dipole approximation. Of course these fields already show a time-dependence due to oscillations at the laser frequency~$\sim e^{-i \omega_L t}$. But such rapid oscillations cancel in the rotating wave approximation and thus a time-dependent term $\vct{B}\times\vct{d}$ can only arise if either the laser field and/or the atomic dipole are modulated separately at a slower time--scale.

The usual interaction between a laser of frequency~$\omega_L$ and an atom gives rise to forces $F \propto \hbar \omega_L/c$. If this laser field is modulated in amplitude or phase on a time--scale~$\sigma t$ we will see that~$\tder{t}{} \big[\vct{B}\times\vct{d}\big]$ gives rise to an additional force $f\propto \hbar \sigma/c$. For technical reasons and to avoid conflicts with the rotating wave approximation it is beneficial to assume that any phase-- or amplitude modulation of the laser beam will be much slower than the main (optical) frequency,~$\sigma \ll \omega_L$.
%
%
\section{Time-modulated fields interacting with a two-level atom}\label{sec:_2lvlexamples}

Most features arising from the R\"{o}ntgen interaction become apparent in the most simple setup of a two-level atom interacting with a single, amplitude-modulated laser beam. Although this example has been partially discussed before~\cite{hinds2009momentum}, it is a useful foundation for the more developed setup discussed later in Sect.~\ref{sec:_4lvl_examples}. 

For a suitably chosen interaction picture and in the rotating-wave approximation, the Hamiltonian describing a two-level atom interacting with an electromagnetic field is
\begin{equation}
	H=\frac{\vct{P}^2}{2M} - \hbar \delta \ketbra{e}{e} + H_\text{AL} + H_\text{AV}	\,,
\end{equation}
where~$\delta=\omega_L-\omega_A$ describes the detuning between the laser and the atomic transition frequency. The coupling to the classical laser field propagating in a direction~$\bkappa = \bk c/\omega_L$ is given by
\begin{align}
	H_\text{AL} &= - \vct{d}\cdot\vct{E}_L - \tfrac{1}{2M}\big[\vct{P} \cdot (\vct{B}_L\times \vct{d}) + (\vct{B}_L\times \vct{d}) \cdot \vct{P} \big]
				\notag \\
		&= - \big(1 - \tfrac{1}{M c} \vct{P}\cdot \bkappa + \tfrac{\hbar \omega_L}{2 M c^2} \big) \vct{d}^{(-)}\cdot \vct{E}_L^{(+)}
			- \tfrac{1}{M c} (\vct{P} \cdot \vct{E}_L^{(+)}) ( \vct{d}^{(-)} \cdot \bkappa)
			+ \Hc \,, \label{eq:_Hatomlaser}
\end{align}
with the laser field $\vct{E}_L^{(\pm)} = \tfrac{1}{2} \bepsilon \mathcal{E}(\vct{R},t) \exp(\pm i \bk\cdot \vct{R})$ and the dipole operators for a two-level atom with states~$\ket{g}$ and~$\ket{e}$, $\vct{d}^{(-)} = \bkewreduced{g}{\vct{d}}{e} \ketbra{e}{g}$ and $\vct{d}^{(+)}= \bkewreduced{g}{\vct{d}}{e} \ketbra{g}{e}$ where $\bkewreduced{g}{\vct{d}}{e}$ is the corresponding reduced dipole matrix element. To arrive at the second line we used the fact that $\vct{B}_L=\bkappa \times \vct{E}_L/c$, the commutation relation $\exp(\pm i \bk \cdot \vct{R})\vct{P} = (\vct{P}\mp \hbar \bk) \exp(\pm i \bk \cdot \vct{R})$ and the ``bac-cab rule'' for cross-products. The fast time-dependence of the laser fields~$\sim\exp(\pm i \omega_L t)$ has been absorbed into the dipole operators during the rotating wave approximation.

The interaction between the atom and the vacuum is described by~$H_\text{AV}$ which has a form similar to~$H_\text{AL}$. The effects of the vacuum Hamiltonian will only be included phenomenologically as spontaneous decay rates when we discuss the evolution of the atomic states during the interaction with the light field.

Throughout this work we assume a well localised atom in a semi-classical approximation such that we can replace the position and momentum operators~$\vct{R}$ and~$\vct{P}$ by their expectation values~$\vct{r}$ and~$\vct{p}$~\cite{cohen1992atom}.

The basic features of the R\"{o}ntgen-term in connection with time-dependent optical fields can be discussed using an amplitude-modulated plane wave propagating along the $+z$-direction, \ie $\bkappa=\be_z$. In this case we simplify the amplitude function $\mathcal{E}(\vct{R},t) \rightarrow \mathcal{E}(\zeta)$, where $\zeta:= \sigma t - \sigma z/c$ is dimensionless and $\sigma \ll \omega_L$ sets the time scale of this modulation. A field propagating in the $-z$-direction, \ie $\bkappa=-\be_z$, would evolve as~$\sigma t + \sigma z/c$, \cf Sect.~\ref{sec:_Ex_sigmaPM_and_Bfield}.

As usual for effective two-level systems, the average direction of the dipole, $\ew{\vct{d}}$ is parallel to the polarisation of the electric field such that $\vct{d}^{(-)} \cdot \bkappa = 0$ as $\bepsilon \perp \bkappa$. Neglecting also the small recoil term~$\hbar\omega_L/(2Mc^2)$, the atom-laser interaction from Eq.~\eqref{eq:_Hatomlaser} can then be written as
\begin{equation}\label{eq:_Hatomlaser_2lvlatom}
	H_\text{AL} = \tfrac{1}{2} (1- p_z/Mc) \hbar \Omega_L(\zeta) \left( \mathcal{S}_{e g} e^{i k z} + \mathcal{S}_{g e} e^{-i k z} \right) \,,
\end{equation}
where $\mathcal{S}_{a b} : = \ketbra{a}{b}$ and $\hbar \Omega_L(\zeta) := - \bkewreduced{g}{\vct{d}\cdot\bepsilon}{e} \mathcal{E}(\zeta)$.

The equations of motion in this semi-classical treatment are given by $\dot{\br}=\ew{\tfrac{i}{\hbar} [H,\vct{R}]} =\ew{\tpder{\bp}{}H}$ and $\dot{\bp}=-\ew{\tpder{\br}{}H}$, such that
\begin{subequations}\label{eq:_EquOfMotion_singlemode}
\begin{align}
	M\dot{z}	&=	p_z - \tfrac{\hbar}{c} \Omega_L(\zeta) u(z,t) \,,
		\label{eq:_EquOfMotion_singlemode_zdot}
\\	
	\dot{p}_z	&=	\hbar \big(1- \tfrac{p_z}{Mc}\big) \big[ k \Omega_L(\zeta) v - u \tpder{z}{} \Omega_L(\zeta) \big] \,,
		\label{eq:_EquOfMotion_singlemode_pzdot}
\end{align}\end{subequations}
where $u(z,t)$ and $v(z,t)$ are the real and imaginary parts of $\ew{\mathcal{S}_{e g}} e^{i k z}$ and are proportional to the coherences in the atom's density matrix. Note the clear difference between the canonical and the kinetic momentum as outlined in Eq.~\eqref{eq:_kinetic_canonic_momentum}.

Aside from a factor $(1- p_z/Mc)$ we recognise that the first component in $\dot{p}_z$ describes the radiation pressure pushing the particle along the direction of beam propagation~\cite{cohen1992atom}. The second part corresponds to the the gradient- or dipole force accelerating the particle along the gradient of the electric field intensity. Here this gradient is not due to the focussing of the beam or interference effects forming a standing wave pattern, but due to the time-modulation of the amplitude,
\begin{equation}\label{eq:_Omega_L_derivatives}
	\tpder{z}{} \Omega_L(\zeta) = - \tfrac{1}{c}\tpder{t}{} \Omega_L(\zeta) = -\tfrac{\sigma}{c} \Omega_L'(\zeta)	\,,
\end{equation}
with $\Omega_L'(\zeta) := \tder{\zeta}{} \Omega_L(\zeta)$. In Eq.~\eqref{eq:_EquOfMotion_singlemode_pzdot} we thus have a radiation-pressure component~${\sim \hbar k \Omega_L}$ and a gradient-force component~${\sim \hbar \tfrac{\sigma}{c} \Omega_L'}$. For a red detuned laser beam, \ie for $\delta = \omega_L-\omega_A <0$, we will see in Eq.~\eqref{eq:_OBE_singlemode_approxsol} that $u$ is negative and we expect the gradient force to drag particles towards the travelling maximum of~$\Omega_L(\zeta)$.

It is important to notice that Eq.~\eqref{eq:_EquOfMotion_singlemode_pzdot} only gives the change of the canonical momentum. The change in kinetic momentum is given by a time-derivative of Eq.~\eqref{eq:_EquOfMotion_singlemode_zdot},
\begin{align}
	M \ddot{z} &= \dot{p}_z - \tfrac{\hbar}{c} \tder{t}{} \big[ \Omega_L(\zeta) u(z,t)\big]	\notag \\
		&= \hbar (1- \beta_z) \big[ k \Omega_L v - u \tpder{z}{} \Omega_L \big]
			- \tfrac{\hbar}{c} \Big(
			\big[ \big( \tpder{t}{} + c \beta_z \tpder{z}{} \big) \Omega_L\big] u
			+ \Omega_L \dot{u}
			\Big) \label{eq:_Force_singlemode_intermediate}
\end{align}
where we used Eq.~\eqref{eq:_EquOfMotion_singlemode_pzdot}, set $\tder{t}{}=\tpder{t}{}+c \beta_z \tpder{z}{}$ and  identified $\frac{p_z}{Mc} \approx \dot{z}/c =:\beta_z$ by dropping terms of order~$\frac{\hbar \Omega_L}{Mc^2}$~\footnote{Of course we ignored terms associated with the change in mass-energy as we also dropped the recoil terms~$\sim \hbar \omega_L/(2 Mc^2)$~\cite{sonnleitner2017will}.}. Using the relations from Eq.~\eqref{eq:_Omega_L_derivatives} we thus get a force in the $z$-direction of the form
\begin{equation}\label{eq:_Force_singlemode_intermediate2}
	M \ddot{z} = \hbar (1- \beta_z) k \Omega_L(\zeta) v(z,t) - \tfrac{\hbar}{c} \Omega_L(\zeta) \dot{u}(z,t)	\,.
\end{equation}

As usual, the evolution of the atomic states and the associated quantities $u$ and $v$ is calculated by first deriving the Heisenberg equations of the atomic operators,$\tder{t}{} \mathcal{S}_{a b} = \tfrac{i}{\hbar} [H, \mathcal{S}_{a b}]$ and then solving the optical Bloch equations
\begin{subequations}\label{eq:_OBE_singlemode}
\begin{align}
	\dot{u} &=   \left(\delta - \omega_L \beta_z \right) v - \tfrac{\Gamma}{2} u\,,\\
	\dot{v} &=	- \left(\delta - \omega_L \beta_z \right) u - (1- \beta_z) \Omega_L(\zeta) w -\tfrac{\Gamma}{2} v \,,\\
	\dot{w} &=	(1- \beta_z) \Omega_L(\zeta) v -\Gamma \big( w + \tfrac{1}{2} \big)	\,,
\end{align}
\end{subequations}
where $w:=\ew{S_{e e}-S_{g g}}/2$. For slowly varying fields, $\sigma \ll \Gamma$, these equations can be solved using an adiabatic approximation such that we get, in the low-saturation regime~\cite{cohen1992atom},
\begin{subequations}\label{eq:_OBE_singlemode_approxsol}
\begin{align}
	u(\zeta) &\approx (\delta - \omega_L \beta_z) \frac{(1- \beta_z) \Omega(\zeta)/2}{(\delta - \omega_L \beta_z)^2+\Gamma^2/4} \,,
	\\
	v(\zeta) &\approx \frac{\Gamma}{2} \frac{(1- \beta_z) \Omega(\zeta)/2}{(\delta - \omega_L \beta_z)^2+\Gamma^2/4} \,,
	\\
	\dot{u}(\zeta) &= \sigma (1-\beta_z) u'(\zeta) \approx \sigma (1-\beta_z) \frac{\Omega'(\zeta)}{\Omega(\zeta)} u(\zeta) \,.
\end{align}
\end{subequations}
If the field changes on time scales comparable to or greater than the decay rate ($\sigma \gtrsim \Gamma$), then the (typically numerical) solutions of the optical Bloch equations start to show non-adiabatic behaviour. 

\begin{figure}
	\centering
	\includegraphics[width=\textwidth]{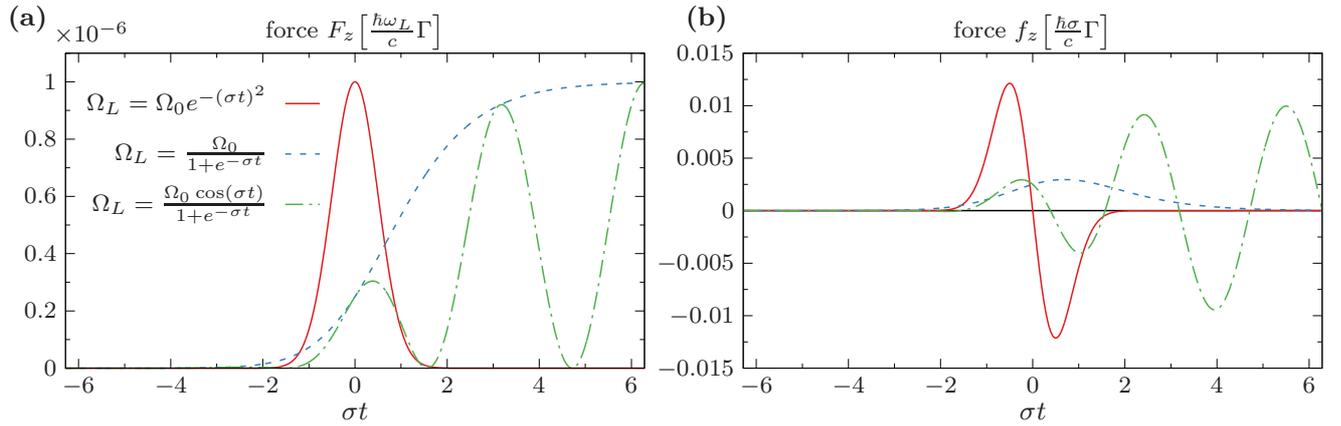}
	\caption{\label{fig:_2lvl_atom_singlebeam}%
	The forces experienced by a two-level atom at rest interacting with time-varying single laser beam for different shapes of the amplitude modulation ($\Omega_L = \Omega_0 \exp[-(\sigma t)^2/2]$, red, solid line; $\Omega_L=\Omega_0/[1+\exp(-\sigma t)]$, blue, dashed line; $\Omega_L=\Omega_0 \cos(\sigma t)/[1+\exp(-\sigma t)]$, green, dash-dotted curve). Figure~(a) shows the usual radiation pressure force~$F_z$ (first term in Eq.~\eqref{eq:_Force_singlemode}) as the amplitude modulation passes the atom; this force follows the temporal shape of the beam intensity as experienced by the atom. Figure~(b) shows the force~$f_z$, the second term in Eq.~\eqref{eq:_Force_singlemode}; this force is proportional to the derivative of the field, but it is positive (negative) for an increasing (decreasing) intensity, which is exactly the opposite of what one would expect from the dipole force for negative detuning. Note the different units for the two forces as $\sigma/\omega_L \ll 1$. Parameters used here are $\delta/\Gamma=-10^4$, $\Omega_0/\Gamma=20$, $\sigma/\Gamma=1/100$.}
\end{figure}

Using the approximate solutions from above, we obtain for the total force
\begin{equation}\label{eq:_Force_singlemode}
		M \ddot{z} \approx (1- \beta_z) \left( \tfrac{\hbar \omega_L}{c} \Omega_L(\zeta) v(\zeta) - \tfrac{\hbar \sigma}{c} \Omega_L'(\zeta) u(\zeta) \right)	\,
\end{equation}
When we compare this result to what we had in Eq.~\eqref{eq:_EquOfMotion_singlemode_pzdot} we see that the gradient force~${\propto - u(\zeta) \partial_z \Omega_L(\zeta) }= u(\zeta) \tfrac{\sigma}{c} \Omega_L'(\zeta)$ has been cancelled out by an equal, but opposite term arising from Eq.~\eqref{eq:_EquOfMotion_singlemode_zdot}, \cf also Eq.~\eqref{eq:_Force_singlemode_intermediate}. The (approximate) solution of the optical Bloch equations then gives rise to a new term proportional to the gradient of $\Omega_L$, \emph{but of opposite sign}~\cite{hinds2009momentum}. As shown in Fig.~\ref{fig:_2lvl_atom_singlebeam}, a red detuned laser will drag the atom to the \emph{minimum} of $\Omega_L$. We thus see that the presence of the R\"{o}ntgen term and the associated difference between the canonical and kinetic momentum reverts the direction of dipole-acceleration due to the gradient associated with the time-modulation of an em-field.

Of course, this unexpected behaviour only affects the gradient force due to the time-modulation propagating with~$\sigma(t \mp z/c)$. Gradient forces arising from a stationary setup, \ie from a focussed beam or interference effects, behave as usual.

Generalising from the example of Eq.~\eqref{eq:_Force_singlemode} we note that a ubiquitous feature of forces on atoms in time-modulated fields is that they can be split in two parts: One component is proportional to $\hbar \omega_L/c$ and contains the well known forces as they also arise in stationary fields; aside from factors~${\sim(1-\beta_z)}$ these forces remain unchanged by the presence of the R\"{o}ntgen term. The second component contains the time-derivatives of the modulated quantities (here $\Omega_L(\zeta)$) and related changes in the internal atomic evolution (here given by $\dot{u}$); this component is proportional to $\hbar \sigma/c$ and is usually very weak as $\sigma \ll \omega_L$ by design.

In the following discussions we shall assume that the atom is at rest, $\beta_z=0$, and split the total force in two parts $M \ddot{\br} = \mathbf{F} + \mathbf{f}$, where $\mathbf{F}\propto \hbar \omega_L/c$ and $\mathbf{f} \propto \hbar \sigma/c$. From equs.~\eqref{eq:_Force_singlemode_intermediate2} and~\eqref{eq:_Force_singlemode} we see that $F_z=\hbar \tfrac{\omega_L}{c} \Omega_L(\zeta) v(\zeta)$ while $f_z = {- \frac{\hbar}{c} \Omega_L(\zeta) \dot{u}(z,t)} \approx  {- \hbar \tfrac{\sigma}{c} \Omega_L'(\zeta) u(\zeta)}$ in this example. Figure~\ref{fig:_2lvl_atom_singlebeam} shows these forces for a two-level atom interacting with various forms of amplitude-modulated laser beams. We see that the dominant force contribution~$F_z$ behaves as expected as it pushes particles along the beam with a magnitude directly proportional to the intensity of the field~$\Omega_L^2(\zeta)$. The term~$f_z$ associated with the R\"{o}ntgen-interaction shows the counter-intuitive behaviour discussed above as we get a positive force where $\Omega_L^2$ increases such that the atom is pushed away from the field maximum despite the red-detuned laser frequency. 

The simple form of the force in Eq.~\eqref{eq:_Force_singlemode} derived from the adiabatic solutions~\eqref{eq:_OBE_singlemode_approxsol} allows us to compute the total change in velocity due to the additional force term~$f_z$ as $M (\dot{z}(t_1) - \dot{z}(t_0)) = \int_{t_0}^{t_1} f_z(t) \dd t$,
\begin{equation}
	\int_{t_0}^{t_1} f_z(t) \dd t
		\approx - \frac{\hbar \delta}{4 M c} \frac{\Omega_L^2(\sigma t_1) - \Omega_L^2(\sigma t_0)}{\delta^2 + \Gamma^2/4}	\,.
\end{equation}
We can thus see that, at least for this example, the total effect of the additional force is proportional to the difference in beam intensity at times~$t_0$ and~$t_1$, but is independent of~$\sigma$. However, this net effect remains several orders of magnitude below the recoil velocity for $v_\text{rec}=\hbar \omega_A/(M c)$ as $\delta \ll \omega_A$.

The forces shown in Fig.~\ref{fig:_2lvl_atom_singlebeam} and also later in Figs.~\ref{fig:_fourlvl_singleLaserB_Ex1}, \ref{fig:_fourlvl_singleLaserB_Ex2} and~\ref{fig:4lvl_sigmaPM_Ex2} are given in units $F\sim [\hbar \omega_L \Gamma/c]$ and $f\sim [\hbar \sigma \Gamma/c]$, respectively. For typical optical transitions where $\omega_L\approx 10^{15} \,\mathrm{s}^{-1}$ and $\Gamma \approx 10^{7}\,\mathrm{s}^{-1}$ we have $\Gamma/\omega_L\approx 10^{-8}$, which helps us to relate the scales for forces~$f$ and~$F$ as
\begin{equation}
	\Big[\frac{\hbar \sigma}{c} \Gamma \Big] \approx \frac{\sigma}{\Gamma} 10^{-8} \Big[\frac{\hbar \omega_L}{c} \Gamma	\Big]\,,
\end{equation}
so that usually $f\ll F$.

A more systematic discussion on the magnitude of the R\"{o}ntgen forces is given in Sect.~\ref{sec:_Discussion_HowSmall}.
%
%
\section{Setups involving multiple transitions}\label{sec:_4lvl_examples}

In the previous section we discussed some basic properties of an (induced) electric dipole interacting with a time-modulated field and the  resulting corrections to forces acting along the beam propagation axis. This discussion was based on an effective two-level atom and made use of the first term~$\sim \big(1 - \vct{p}\cdot \bkappa/Mc\big) \vct{d}\cdot \vct{E}_L$ in the interaction Hamiltonian as given in Eq.~\eqref{eq:_Hatomlaser}. 

The second term~$\sim (\vct{p} \cdot \vct{E}_L) ( \vct{d} \cdot \bkappa)$ vanished for $\vct{d} \parallel  \vct{E}_L \leftrightarrow \vct{d} \perp \bkappa$. But we see that if we manage to ``rotate'' the average dipole moment away from the electric field, such that $\ew{\vct{d} \cdot \bkappa}\neq 0$, then we get an interaction~$\sim \vct{p} \cdot \vct{E}_L$ which includes terms proportional to $p_x$ and $p_y$ for a beam propagating along the $z$-axis. As $\dot{\vct{r}} = \ew{\tpder{\bp}{} H}$ we see that these terms give rise to forces perpendicular to the direction of beam propagation, even if the beam is a plane wave with no transverse structure.

\begin{figure}
	\centering
	\includegraphics[width=0.5 \textwidth]{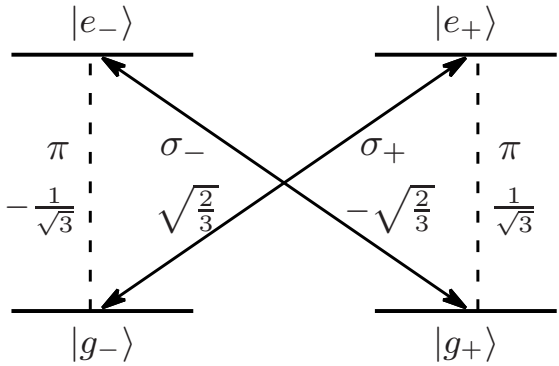}
	\caption{\label{fig:_4lvlsetup}%
	The four-level configuration used in the examples in Sect.~\ref{sec:_4lvl_examples}. Laser beams can induce $\ket{g_-}\leftrightarrow\ket{e_+}$ ($\ket{g_+}\leftrightarrow\ket{e_-}$) transitions if they are $\sigma_+$ ($\sigma_-$) circularly polarised. An additional interaction with an external magnetic field couples the ground states $\ket{g_-}$ and $\ket{g_+}$, \cf Eq.~\eqref{eq:_HB}. The numbers give the Clebsch-Gordan coefficients of the respective transitions as used in Appendix~\ref{appendix:_atomlaserH_multilevel}.}
\end{figure}%
To explore this unexpected possibility we move from an effective two-level system to a four-level system in a $J_g=1/2 \leftrightarrow J_e=1/2$ transition with two ground states with magnetic quantum numbers $m_g=\pm 1/2$ and two degenerate excited states with $m_e=\pm 1/2$ labelled $\ket{g_-}$, $\ket{g_+}$ and $\ket{e_-}$, $\ket{e_+}$, respectively. This configuration is shown in Fig.~\ref{fig:_4lvlsetup} and more details about the dipole operator and the atom-field Hamiltonian for multi-level configurations are given in appendix~\ref{appendix:_atomlaserH_multilevel}. From there we see that the average dipole moment in this setup is
\begin{equation}\label{eq:_average_dipolemoment}
	\ewbig{\vct{d}} \propto
		\ewbig{ \mathcal{S}_{e_+ g_-} + \mathcal{S}_{e_- g_+} } \be_x
		+ i \ewbig{ \mathcal{S}_{e_+ g_-} - \mathcal{S}_{e_- g_+} } \be_y
		+ \ewbig{ \mathcal{S}_{e_+ g_+} - \mathcal{S}_{e_- g_-} } \be_z
		+ \cc \,,
\end{equation}
where $\mathcal{S}_{e_+ g_-} := \ketbra{e_+}{g_-}$, \etc

Equation~\eqref{eq:_average_dipolemoment} shows that $\ew{\vct{d} \cdot \bkappa} \propto \ewbig{ \mathcal{S}_{e_+ g_+} - \mathcal{S}_{e_- g_-}} + \cc$ for a laser beam propagating along the quantisation axis, $\bkappa \parallel \be_z$. Such a laser beam will usually only drive $\sigma_\pm$-transitions coupling $\ket{g_-}\leftrightarrow \ket{e_+}$ and $\ket{g_+}\leftrightarrow \ket{e_-}$, respectively~\footnote{The notation for $\sigma_\pm$-polarised beams should not be confused with the time-scale of beam modulation~$\sigma$.}. ``Rotating the dipole'' thus requires a mechanism that allows for coherent $\pi$-transitions ($\ket{g_-}\leftrightarrow \ket{e_-}$ and $\ket{g_+}\leftrightarrow \ket{e_+}$) such that $\ewbig{d_z}\neq 0$.

In the examples discussed below this coupling is induced by an additional isotropic magnetic field along the $y$-axis driving transitions $\ket{g_-}\leftrightarrow\ket{g_+}$ with a Larmor frequency $\Omega_B = \gamma B$, with $\gamma$ being the gyromagnetic ratio of the ground state. In the Hamiltonian we thus add a term
\begin{equation}\label{eq:_HB}
	H_\text{B}= - i \tfrac{\hbar}{2} \Omega_B \mathcal{S}_{g_- g_+} + \Hc \,.
\end{equation}
Here we dropped the corresponding term for Larmor transitions within the excited states, $\ket{e_-}\leftrightarrow\ket{e_+}$, as it would only increase our parameter space without substantially changing the dynamics.

Of course, this, possibly time-dependent, magnetic field would also couple to the electric dipole of the atom via the R\"{o}ntgen interaction or via the electric field arising from $\partial_t \vct{B} = - \nabla \times \vct{E}$. But these terms vanish in the rotating wave approximation where the dipole follows the laser field and oscillates with the laser frequency.

We shall also assume that the coupling to the external magnetic field is isotropic, $\nabla \Omega_B = 0$, but might change in time, $\partial_t \Omega_B\neq 0$. This is the case near the centre of a pair of Helmholtz coils where anisotropies due to retardation effects can be ignored if the driving current is changed simultaneously for both coils. 

In the following examples we shall first discuss the forces arising from the combination of a single $\sigma_+$-polarised laser beam and a magnetic field. In the second example we present results from combining this magnetic field with counter-propagating $\sigma_+$ and $\sigma_-$ beams of slightly different detuning.

\subsection{Example: A four-level configuration, a magnet and a single laser}\label{sec:_Ex_sigmaP_and_Bfield}

The setup for this example is inspired by the work by R.~Kaiser~\etal who used it to discuss their theory and experiment on the mechanical Hanle effect~\cite{kaiser1991mechanical}. Considering the level structure shown in Fig.~\ref{fig:_4lvlsetup} we see that a $\sigma_+$ circularly polarised laser beam will drive the $\ket{g_-}\leftrightarrow\ket{e_+}$ transition until spontaneous decay from $\ket{e_+}$ to $\ket{g_+}$ traps the atomic state in $\ket{g_+}$, which does not interact with the laser any further. In the steady-state limit the resulting force from the laser will thus vanish because the $\ket{g_-}$--state is empty.

But adding a magnetic interaction~$H_\text{B}$ as given in Eq.~\eqref{eq:_HB} enables a closed loop $\ket{g_-}\leftrightarrow\ket{e_+}\rightarrow\ket{g_+} \leftrightarrow \ket{g_-}$ with a continuous radiation pressure force along the beam axis. R.~Kaiser~\etal used this mechanism to measure small magnetic fields and their effect on the deflection of Helium atoms traversing a laser beam~\cite{kaiser1991mechanical}. 

Let us consider this example more closely, include the Röntgen-term and allow for time-dependent fields. Similar to the previous examples we describe the $\sigma_+$-polarised laser beam travelling in the $+z$-direction as $\mathbf{E}_L = \tfrac{1}{2} \mathcal{E}(\zeta) \be_{-1}^\ast  \exp{[i (kz - \omega_L t)]} + \text{c.c.}$ and define $\hbar \Omega_L(\zeta):= - \sqrt{2/3} \bkewreduced{J_g}{\vct{d}}{J_e}\mathcal{E}(\zeta)$, where $\bkewreduced{J_g}{\vct{d}}{J_e}$ is the reduced dipole matrix element of this transition, \cf Appendix~\ref{appendix:_atomlaserH_multilevel}, Eq.~\eqref{eq:_dipoleOP_4lvl_0comp}. We thus get a Hamiltonian $H=\bp^2/(2M) - \hbar \delta \big( \mathcal{S}_{e_+ e_+} + \mathcal{S}_{e_- e_-} \big) + H_\text{B} + H_\text{AL} + H_\text{AV}$ where
\begin{equation}\label{eq:_Hatomlaser_4lvlatom}
	H_\text{AL}= \frac{\hbar \Omega_L(\zeta)}{2} \Big(1 - \frac{p_z}{Mc}\Big) \mathcal{S}_{e_+ g_-} e^{i k z} 
	+ \frac{\hbar \Omega_L(\zeta)}{2} \frac{p_x + i p_y}{2Mc} \left( \mathcal{S}_{e_+ g_+} - \mathcal{S}_{e_- g_-} \right) e^{i k z} + \Hc \,,
\end{equation}
and $H_\text{B}$ is defined in Eq.~\eqref{eq:_HB}. The corresponding equations of motion then give $\dot{p}_x= \dot{p}_y=0$ and
\begin{subequations}\label{eq:_EquOfMotion_Example3}
\begin{align}
	M \dot{x} &= p_x
		+ \tfrac{\hbar}{2 c} \Omega_L(\zeta) \big( u_{e_+ g_+} - u_{e_- g_-} \big)	\,, \\
	M \dot{y} &= p_y 
		- \tfrac{\hbar}{2 c} \Omega_L(\zeta) \big( v_{e_+ g_+} - v_{e_- g_-} \big)	\,, \\
	M \dot{z} &= p_z
		- \tfrac{\hbar}{c} \Omega_L(\zeta) u_{e_+ g_-} \,, \\
	\dot{p}_z &= \hbar k \Omega_L(\zeta) v_{e_+ g_-}
	+ \hbar \tfrac{\sigma}{c} \Omega_L'(\zeta) u_{e_+ g_-}
	\,,
\end{align}
\end{subequations}
where we dropped terms~$\sim \mathcal{O}(\tfrac{p_{x,y,z}}{Mc})$ and used $u_{e_+ g_+}(z,t) := \tfrac{1}{2} \ew{ \mathcal{S}_{e_+ g_+} e^{ikz} + \Hc}$, $v_{e_+ g_+}(z,t) := \tfrac{1}{2i} \ew{ \mathcal{S}_{e_+ g_+} e^{ikz} - \Hc}$, etc. These are solutions to the in total 15 optical Bloch equations describing the evolution of the atomic states given in Appendix~\ref{appendix:_OptBloch_Ex1}.

\begin{figure}
	\centering
	\includegraphics[width=0.5 \textwidth]{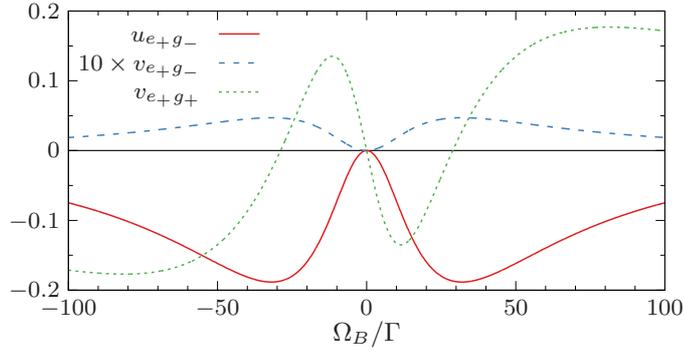}
	\caption{\label{fig:_fourlvl_singleLaserB_SS_uv}%
	The steady state solutions for the atomic states of the four-level configuration shown in Fig.~\ref{fig:_4lvlsetup} as defined below Eq.~\eqref{eq:_EquOfMotion_Example3} (see also appendix~\ref{appendix:_OptBloch_Ex1}) plotted as a function of the magnetic interaction strength~$\Omega_B/\Gamma$, \cf Eq.~\eqref{eq:_HB}, for a constant laser field, $\Omega_L/\Gamma=50$, and a detuning $\delta/\Gamma=-20$.  One clearly sees how $u_{e_+ g_-}$ (red, solid line), $v_{e_+ g_-}$ (blue, dashed line) and $v_{e_+ g_+}$ (green, dotted line) change strongly in the presence of weak magnetic fields. The quantities not shown here, $u_{e_+ g_+}$, $u_{e_- g_+}$, $v_{e_- g_+},$ \etc are zero for all values of $\Omega_B$. Although these are solutions for a stationary setup, they are useful to estimate the behaviour for time-dependent laser- or magnetic fields shown in Figs.~\ref{fig:_fourlvl_singleLaserB_Ex1} and~\ref{fig:_fourlvl_singleLaserB_Ex2}, respectively.}
\end{figure}%
Fig.~\ref{fig:_fourlvl_singleLaserB_SS_uv} shows the corresponding steady state solutions for constant~$\Omega_L$ as a function of $\Omega_B$ (which is also constant). We see that $u_{e_+ g_-}$, $v_{e_+ g_-}$ and $v_{e_+ g_+}$ vanish for $\Omega_B=0$ and that the dominant force $\vct{F} = \hbar k \Omega_L v_{e_+ g_-} \be_z$ measured in Ref.~\cite{kaiser1991mechanical} changes significantly in the presence of a weak magnetic coupling~$\Omega_B$.

From the equations of motion given in Eq.~\eqref{eq:_EquOfMotion_Example3} we can derive all terms arising from the Röntgen-interaction or directly from~$\Omega_L'(\zeta)$ to generate the force~$\vct{f}\sim\hbar\sigma/c$,
\begin{subequations}\label{eq:_weakForce_Example3}
\begin{align}
	f_x &= \tfrac{\hbar \sigma}{2 c} \Omega_L'(\zeta) \big( u_{e_+ g_+} - u_{e_- g_-} \big)
			\notag \\ &\qquad
			+ \tfrac{\hbar}{2 c} \Omega_L(\zeta) \big( \dot{u}_{e_+ g_+} - \dot{u}_{e_- g_-} \big) 	\,, \\
	f_y &= - \tfrac{\hbar \sigma}{2 c} \Omega_L'(\zeta) \big( v_{e_+ g_+} - v_{e_- g_-} \big)	
			\notag \\ &\qquad
			- \tfrac{\hbar}{2 c} \Omega_L(\zeta) \big( \dot{v}_{e_+ g_+} - \dot{v}_{e_- g_-} \big) 	\,, \\
	f_z &= - \tfrac{\hbar}{c} \Omega_L(\zeta) \dot{u}_{e_+ g_-}	\,.
\end{align}
\end{subequations}
In Fig.~\ref{fig:_fourlvl_singleLaserB_SS_uv} we see that the steady state solution of $v_{e_+ g_+}$ is also non-zero and varies even more strongly than $v_{e_+ g_-}$ for~$\Omega_B\neq0$. We therefore expect force components in the~$x$ or~$y$ direction in the presence of time-modulated fields~$\Omega_L$ or~$\Omega_B$. Here, we again assume an atom at rest which also simplifies the (numerical) solution of the optical Bloch equations.

\begin{figure}
	\centering
	\includegraphics[width=\textwidth]{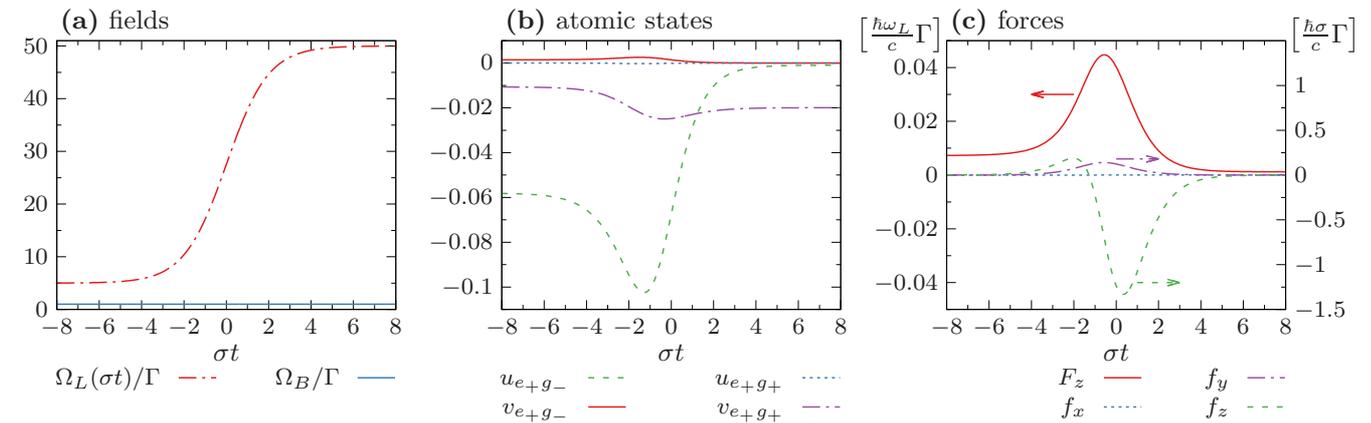}
	\caption{\label{fig:_fourlvl_singleLaserB_Ex1}%
	Forces due to a combination of a time-modulated laser field and a constant magnetic field. Panel~(a) shows the constant magnetic field $\Omega_B/\Gamma=1$ and a strongly increased laser field $\Omega_L(\sigma t)$; Fig.~(b) shows the corresponding evolution of the atomic state matrix elements defined below Eq.~\eqref{eq:_EquOfMotion_Example3}, see also appendix~\ref{appendix:_OptBloch_Ex1}. Figure~(c) shows the dominant force component along the beam axis, $F_z$, in units of $[\hbar \omega_L \Gamma /c]$ (red, solid line, left ordinate) and the weaker forces $f_{x,y,z}\sim [\hbar \sigma \Gamma /c]$ (broken lines, right ordinate) as given in Eq.~\eqref{eq:_weakForce_Example3}. Here we see a small component~$f_y$ acting perpendicular to the beam-propagation axis even though the laser is modelled as a plane wave propagating along~$z$. Other parameters are $\delta/\Gamma=-20$ and $\sigma/\Gamma = 0.1.$}
\end{figure}%
\begin{figure}
	\centering
	\includegraphics[width=\textwidth]{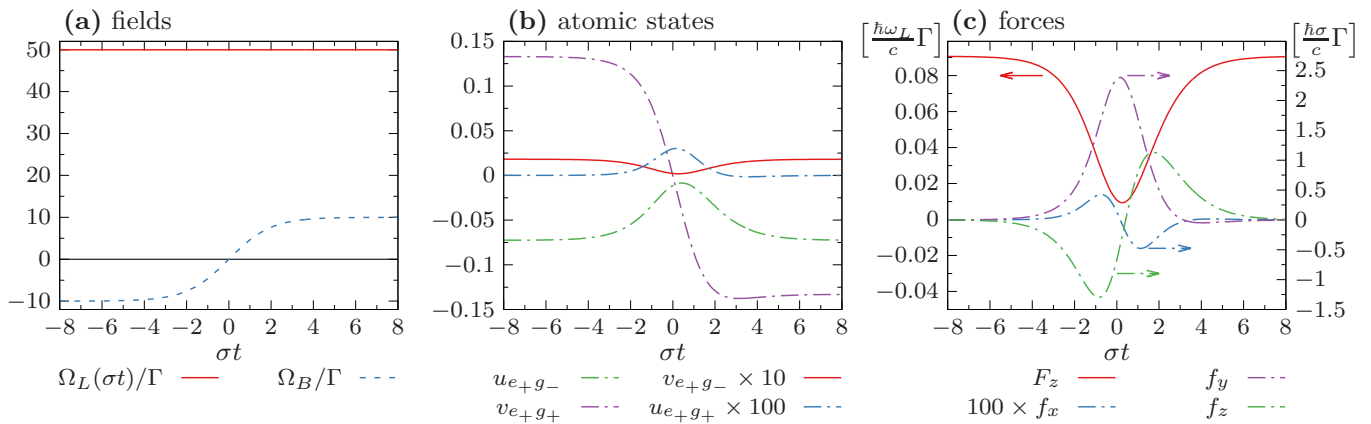}
	\caption{\label{fig:_fourlvl_singleLaserB_Ex2}%
	Forces due to a combination of a constant laser field and a varying magnetic field. Panel~(a) shows the strong constant laser field $\Omega_L/\Gamma= 50$ and the time-dependent Larmor frequency $\Omega_B(\sigma t)$; Fig.~(b) shows the corresponding evolution of the atomic state matrix elements defined below Eq.~\eqref{eq:_EquOfMotion_Example3}. Figure~(c) shows the dominant force component along the beam axis, $F_z$, in units of $[\hbar \omega_L \Gamma /c]$ (red, solid line, left ordinate) and the weaker forces $f_{x,y,z}\sim [\hbar \sigma \Gamma /c]$ (broken lines, right ordinate) as given in Eq.~\eqref{eq:_weakForce_Example3}. As in Fig.~\ref{fig:_fourlvl_singleLaserB_Ex1} we obtain non-vanishing force components acting perpendicular to the beam-propagation axis while $F_z$ drops as $\Omega_B$ crosses zero. Other parameters are $\delta/\Gamma=-20$ and $\sigma/\Gamma = 0.1.$}
\end{figure}%
In Figs.~\ref{fig:_fourlvl_singleLaserB_Ex1} and~\ref{fig:_fourlvl_singleLaserB_Ex2} we show examples for cases where either a strong laser field~$\Omega_L$ or the magnetic field $\Omega_B$ is changed on a time-scale $\sigma t$. The behaviour of both the atomic states and the resulting forces roughly follow what one would expect from the steady-state solutions displayed in Fig.~\ref{fig:_fourlvl_singleLaserB_SS_uv}, although the solutions are not entirely adiabatic because we set $\sigma/\Gamma = 0.1$.

For non-zero magnetic fields we get strong forces~$F_z$ as well as weaker contributions for~$f_z$ or~$f_y$. The component~$f_x$ depends on~$u_{e_+ g_+}$ which vanishes in the steady-state case and is only barely non-zero for the dynamic setups. When we compare the example given in Fig.~\ref{fig:_fourlvl_singleLaserB_Ex2} with the steady-state solutions shown in Fig.~\ref{fig:_fourlvl_singleLaserB_SS_uv} we see that the time-dependentent magnetic interaction shown in Fig.~\ref{fig:_fourlvl_singleLaserB_Ex2}a is chosen to vary such that $v_{e_+ g_+}$ starts at a maximum and ends at its minimum. This ensures a very steep gradient at $\sigma t=0$ and a correspondingly strong force $f_y$ in panel~c.

Let us emphasise that the forces in $y$-direction are not a direct result of the magnetic field oriented along the same axis. This magnetic field is assumed spatially homogeneous, but together with the laser field it enables a closed pumping circle and a coherent coupling between $\ket{e_+}$ and $\ket{g_+}$ (\ie a non-vanishing component $v_{e_+ g_+}$). Intuitively speaking, this magnetic field leads to the rotation of the average electric dipole moment discussed in Eq.~\eqref{eq:_average_dipolemoment}.

\subsection{Example: A four-level configuration, a magnet and two counter-propagating laser beams}\label{sec:_Ex_sigmaPM_and_Bfield}

As a final example we shall extend the setup from above by adding a second, counter-propagating laser beam polarised such that it drives the $\sigma_-$-transition $\ket{g_+}\leftrightarrow \ket{e_-}$, \cf Fig.~\ref{fig:_4lvlsetup}. But in contrast to the previous examples we now assume both laser field amplitudes and the magnetic interaction~$\Omega_B$ to be constant in time. The time-modulation required for the contribution from the Röntgen-term is generated by a phase modulation as one beam shall have a slightly different frequency, $\omega_L'=\omega_L+\sigma$, again with $\sigma \ll \omega_L$.

The resulting Hamiltonian, shown in Eq.~\eqref{eq:_Hexample3_Appendix}, can be constructed straightforwardly by adding up corresponding atom-laser interaction terms~$H_\text{AL}$ as given in Eq.~\eqref{eq:_Hatomlaser} with effective fields
\begin{subequations}\label{eq:_Fields_Example4}
\begin{align}
	\vct{E}_l &= \tfrac{1}{2} \alpha \mathcal{E} \be_{-1}^\ast e^{i kz} e^{-i \sigma (t-z/c)}	+ \cc	\,,\\
	\vct{E}_r &= \tfrac{1}{2} \mathcal{E} \be_{+1}^\ast e^{-i kz} + \cc		\,.
\end{align}
\end{subequations}
The beam $\vct{E}_l$ thus travels in the $+z$-direction, is polarised such that it drives the $\sigma_+$-transition (\cf appendix~\ref{appendix:_atomlaserH_multilevel}) and has a frequency $\omega_L+\sigma$. The beam $\vct{E}_r$ travels in the opposite direction and drives $\sigma_-$-transitions at a frequency~$\omega_L$. The real parameter~$\alpha$ can be used to adjust the relative power of the beams.

Just as in the previous examples it is straightforward to derive the evolution equations for the atomic states, the equations of motion and the corresponding forces $\mathbf{F}\sim \hbar \omega_L \Gamma/c$ and $\mathbf{f}\sim \hbar \sigma \Gamma/c$ from $M \ddot{\br} = \tpder{t}{} \ew{\tpder{\bp}{} H}$,
\begin{subequations}\label{eq:_Forces_Example4}
\begin{align}
	F_z &=	\hbar k \Omega_L (\alpha v_{e_+ g_-} + v_{e_- g_+})	\,,\\
	f_x &=	- \frac{\hbar \Omega_L}{2c} \pder{t}{} \Big[ \big( u_{e_+ g_+}-u_{e_- g_-} \big)\big( \alpha \cos(kz-\zeta) + \cos(kz)\big)
		- \big( v_{e_+ g_+}-v_{e_- g_-} \big)\big( \alpha \sin(kz-\zeta) - \sin(kz)\big) \Big] 
		\,,\\
	f_y &=	\frac{\hbar \Omega_L}{2c} \pder{t}{} \Big[ \big( u_{e_+ g_+}-u_{e_- g_-} \big)\big( \alpha \sin(kz-\zeta) + \sin(kz)\big)
		+ \big( v_{e_+ g_+}-v_{e_- g_-} \big)\big( \alpha \cos(kz-\zeta) - \cos(kz)\big) \Big] 
		\,,\\
	f_z &=	\tfrac{\hbar}{c} \Omega_L \big( \alpha \sigma v_{e_+ g_-} - \tpder{t}{} (\alpha u_{e_+ g_-} + u_{e_- g_+})\big) 	\,.
\end{align}
\end{subequations}

Again we assume an atom initially at rest and set $\zeta = \sigma (t-z/c)$. Here we also set $u_{e_+ g_-}+ i v_{e_+ g_-} := \ew{\mathcal{S}_{e_+ g_-}} \exp[i(kz-\zeta)]$, $u_{e_- g_+}+ i v_{e_- g_+} := \ew{\mathcal{S}_{e_- g_+}} \exp[-i kz]$ while $u_{e_\pm g_\pm}+ i v_{e_\pm g_\pm} := \ew{\mathcal{S}_{e_\pm g_\pm}}$, see appendix~\ref{appendix:_OptBloch_Ex2} for more details.

\begin{figure}
	\centering
	\includegraphics[width=\textwidth]{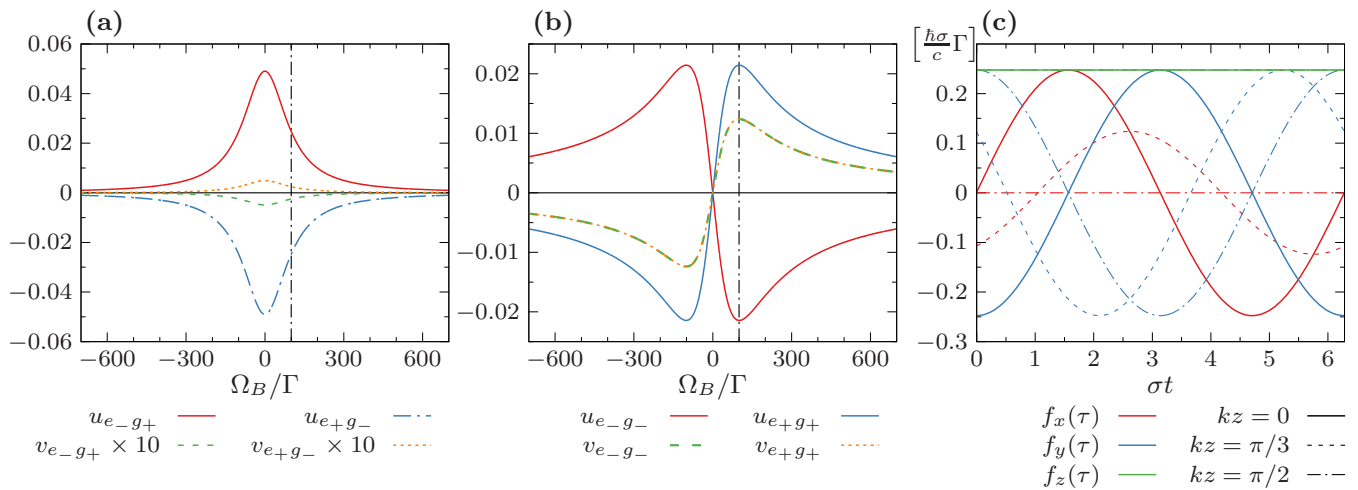}
	\caption{\label{fig:4lvl_sigmaPM_Ex2}%
 	Results for a $\sigma_+$-$\sigma_-$ beam configuration and an additional magnetic field interacting with a $J_g=1/2 \leftrightarrow J_e=1/2$ configuration as shown in Fig.~\ref{fig:_4lvlsetup} and discussed in Sect.~\ref{sec:_Ex_sigmaPM_and_Bfield}. Panels~(a) and~(b) show steady state solutions of the atomic state populations defined below Eq.~\eqref{eq:_Forces_Example4} at position $kz=\pi/3$ when both beams have the same frequency, $\sigma=0$, see also appendix~\ref{appendix:_OptBloch_Ex2}. Due to the large detuning, $\delta/\Gamma =-50$ for $\Omega_L/\Gamma = 10$, we see that $v_{e_\pm g_\mp} \ll u_{e_\pm g_\mp}$. Note that $u_{e_\pm g_\pm}$ and $v_{e_\pm g_\pm}$ vanish in the absence of an external $B$-field and that $v_{e_- g_-}=v_{e_+ g_+}$. The vertical dashed line indicates the value $\Omega_B/\Gamma = 100$ which is used to calculate the time-dependent forces given in Eq.~\eqref{eq:_Forces_Example4}, if the $\sigma_+$-beam has a shifted frequency of $\omega_L+\sigma$ where $\sigma=-2\delta=100 \Gamma$. We see that $f_x$ and $f_y$ oscillate in space and time while $f_z$ is constant, \cf also Fig.~\ref{fig:_4lvl_sigmaPM_Ex2_polarforce}. Here both beams are chosen to have equal power, $\alpha=1$. These parameters are chosen such that the usually dominant force cancels out, $F_z=0$.}
\end{figure}%
Figs.~\ref{fig:4lvl_sigmaPM_Ex2}(a) and~(b) show the steady state solutions ($\sigma=0$, $kz=\pi/3$) of the optical Bloch equations for this system as a function of $\Omega_B/\Gamma$. For~$\sigma \neq 0$ we find that most atomic states are constant in time, only $u_{e_+ g_+}$ and $v_{e_+ g_-}$ can oscillate in time and space. The steady-state solutions are thus again useful to get some intuition for the case~$\sigma \neq 0$. For the detuning chosen in this example, $\delta=-50 \Gamma$, the terms $v_{e_+ g_-}$ and $v_{e_- g_+}$ are strongly suppressed. As they also have opposite sign we see that the usually dominant radiation pressure force cancels in this setup, $F_z \rightarrow 0$~\footnote{Note that in the absence of a magnetic field, $\Omega_B=0$, the two laser beams rely on each other to repopulate the respective ground states. This means that no transition can be stronger than the other and the well known cooling scheme associated with a $\sigma_+$-$\sigma_-$-beam configuration does not work in this case~\cite{dalibard1984potentialities}.}.

\begin{figure}
	\centering
	\includegraphics[width=\textwidth]{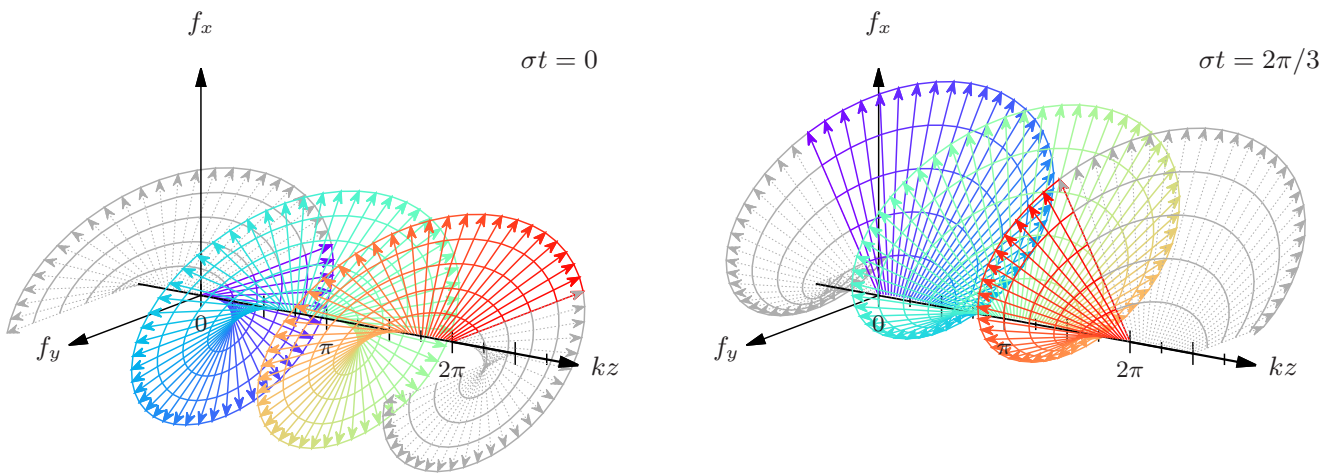}
	\caption{\label{fig:_4lvl_sigmaPM_Ex2_polarforce}%
	Force vector $\vct{f}_\perp=f_x\be_x+f_y\be_y$ for the example given in Fig.~\ref{fig:4lvl_sigmaPM_Ex2} as function of $k z$ at time $\sigma t=0$ (left) and $\sigma t=2 \pi/3$ (right). At $kz=n \pi$, $n \in \mathbb{Z}$, the tip of $\vct{f}_\perp$ follows a circle as $\sigma t$ goes from $0$ to $2\pi$. But this circle continuously changes into a tilted ellipse for other positions until it degenerates into a line at $kz = (n+1)\pi/2$ where $f_y$ oscillates in time while $f_x=0$.}
\end{figure}
From Eq.~\eqref{eq:_Forces_Example4} we see that the components of~$\vct{f}$ depend on the time derivatives of the average atomic states and $\cos(kz-\zeta) = \cos( (\omega_L+\sigma)z/c - \sigma t)$ or $\sin(kz-\zeta)$. We thus expect an oscillating behaviour for $f_x$ and $f_y$ which we also see in Figs.~\ref{fig:4lvl_sigmaPM_Ex2}(c) and~\ref{fig:_4lvl_sigmaPM_Ex2_polarforce}. $f_z$ is constant due to the different photon momenta absorbed from each beam, $\hbar(\omega_L+\sigma)/c$ vs. $\hbar \omega_L/c$. 

The force-components~$f_x$ and~$f_y$ oscillate as functions of $\sigma t$ and $k z$. Their sum, $\vct{f}_\perp=f_x\be_x+f_y\be_y$, spirals along the $z$-axis: atoms at $kz= n \pi$ see a force of constant magnitude which rotates in the $xy$--plane as function of $\sigma t$ while atoms at $kz= (2n+1) \pi/2$ see forces oscillating only along the $y$-direction; at intermediate positions the tip of $\vct{f}_\perp$ follows a tilted ellipse.

The total effect of these oscillating forces obviously averages out when integrated over a period of~$\sigma t$. But experience from both classical and quantum mechanics shows that weak periodic forces combined with harmonic potentials can lead to resonance effects~\cite{meekhof1996generation,lopes2008quantum}. This example might therefore be of special interest in setups where the described counter-propagating beams are combined with a tight radial trap.

\section{Discussion: How small is the R\"{o}ntgen term?}\label{sec:_Discussion_HowSmall}

The examples discussed above show that R\"{o}ntgen forces are intriguing, but also considerably smaller than the usual gradient force and radiation pressure. Here we shall discuss several systematic effects which might lead to forces of a similar magnitude.

As the R\"{o}ntgen term is part of the electric-dipole approximation we have neglected higher order couplings in the atom-light Hamiltonian in Eq.~\eqref{eq:_Hatomlaser}. In comparison to the electric dipole coupling, magnetic dipole or electric quadrupole terms are suppressed by the ratio between the size of the atom (characterised by the Bohr radius) and the wavelength of the laser, $\sim a_0/\lambda$~\cite{steckquantumoptics,cohenQM,loudon2000quantum}. As shown above, the forces resulting from the R\"{o}ntgen term and a time-dependent variation are suppressed with respect to usual dipole forces by a factor~$\sigma/\omega$ which might well be smaller than~$a_0/\lambda$.

Including electric quadrupole interactions is therefore necessary, if the chosen laser frequency is close to a quadrupole-allowed transition. Depending on the frequency and the configuration of atomic energy levels, the forces associated with this transition could be stronger than the R\"{o}ntgen forces discussed here. However, electric quadrupole forces have the same structure as the dominant gradient- or radiation pressure forces as both arise from the spatial derivative of the laser field. R\"{o}ntgen forces, however, arise from the difference between canonical and kinetic momentum as well as a time-derivative of both field and atomic dipole. This is why we focus on the R\"{o}ntgen interaction even if higher-order interaction terms might give stronger effects.

In Sect.~\ref{sec:_4lvl_examples} we showed how R\"{o}ntgen forces can act perpendicular to the beam axis in the presence of an additional magnetic field. As can be seen from the Hamiltonian given in Eq.~\eqref{eq:_HB}, this additional magnetic field can give rise to a force $\mathbf{F}_B = \hbar \nabla \Omega_B v_{g_+ g_-}$. A small inhomogeneity of the magnetic field in the $x$- or $y$-direction can lead to similar perpendicular forces, if $v_{g_+ g_-}$ is non-zero. In the single-beam example the described in Sect.~\ref{sec:_Ex_sigmaP_and_Bfield}, it turns out that $v_{g_+ g_-}$ can be relatively large making this setup sensitive to magnetic field gradients. For the example with counter-propagating beams (\cf Sect.~\ref{sec:_Ex_sigmaPM_and_Bfield}) it turns out that $v_{g_+ g_-}$ vanishes for the chosen setting with $\sigma = -2\delta$.

A transverse gradient from the laser field will also lead to radial forces proportional to $u_{e_+ g_-}$ and $u_{e_- g_+}$. Again, these are non-zero for the single-beam example but are suppressed in the example with counter-propagating beams.

In addition to these systematic effects, stochastic processes such as spontaneous decay can spread the atom's position and momentum uncertainties such that the results of weak forces are concealed. We therefore propose that a measurement of R\"{o}ntgen forces should make use of resonance- or interference phenomena.

%
%
\section{Summary and conclusions}\label{sec:_Conclusion}
In this work we used a semi-classical model to analyse the mechanical interaction between atoms and time-dependent external fields with a special focus on effects due to the Röntgen-term. In Sect.~\ref{sec:_intro_class_discussion} we showed why the Röntgen term and the associated difference between canonical and kinetic momentum is of special interest when radiation fields interacting with an electric dipole are modulated on a time-scale~$\sigma \ll \omega_L$.

In Sect.~\ref{sec:_2lvlexamples} we discussed some of the characteristics of these forces using the example of a simple two-level atom interacting with an amplitude-modulated plane wave. There we could show how the Röntgen-term reverts intuitively expected force terms such that atoms are not pulled towards, but repelled from the travelling intensity maximum of a modulated, red-detuned laser beam.

Using a four-level configuration and an additional external magnetic field to ``rotate the dipole'' we could show how the cross-term~$\vct{d}\times\vct{B}$ in the Röntgen interaction leads to forces perpendicular to the beam propagation axis of a plane wave in Sect.~\ref{sec:_4lvl_examples}. There we also explored the effects of a time-modulated effective dipole axis (achieved by changing the external magnetic field) as well as a configuration involving two slightly detuned laser beams, which effectively corresponds to a time-dependent phase.

These examples made use of laser pulses, time modulated magnetic fields or phase-modulated counter-propagating beams, all of which are ubiquitous in quantum-optical laboratories. Yet they showed that the often ignored Röntgen term opens the way for surprising and counter-intuitive radiation forces. Most importantly, neglecting the Röntgen interaction can lead to wrong results, as demonstrated in Sect.~\ref{sec:_2lvlexamples}.

However, we also discussed that these forces are by design much smaller than usual radiation pressure forces and they might even average out to zero when one considers the net-effect over a longer period. As discussed at the end of Sect.~\ref{sec:_2lvlexamples} and in Sect.~\ref{sec:_Discussion_HowSmall}, the net-effect of these forces alone will usually be many orders of magnitude smaller than that of a single photon recoil or of simple noise effects. But periodic and resonant effect on an otherwise well controlled system, such as on atoms in a harmonic trap, might well be measurable.

The semi-classical calculations presented in this work serve an exploratory purpose, scouting the wealth of phenomena hidden in the Röntgen-term. A quantitative analysis discussing these small effects and their measurability in a specific experimental setup requires a bespoke quantum-mechanical treatment, which will be the focus of future work.

%
%
\section*{Acknowledgements}
We gratefully acknowledge funding by the Austrian Science Fund FWF (J~3703-N27), the EPSRC (QuantIC EP/M01326X/1) and the Royal Society (RP150122).

%
%
\section*{Author Contributions}
All calculations were performed by MS who also devised the examples given in Sect.~\ref{sec:_4lvl_examples}. Both authors contributed to the physical insights and wrote the text.

%
%
\appendix
\section{The atom-laser Hamiltonian for multi-level transitions}\label{appendix:_atomlaserH_multilevel}

As we discuss the atom-laser interaction for a multi-level configuration involving different circularly polarised laser beams in  Sect.~\ref{sec:_4lvl_examples} it is useful to review some concepts and notation. More details can be found in refs.~\cite{steckquantumoptics,cohenQM}.

We use the notation where the fields and dipole operator are described in a spherical basis, which is connected to the Cartesian basis via
\begin{align}
	\be_{\pm 1} &:= \tfrac{1}{\sqrt{2}} ( \mp \be_x - i \be_y)	\,, &
	\be_0 &:= \be_z	\,,
\end{align}
such that $\be_x = -(\be_{1}-\be_{-1})/\sqrt{2}$ and $\be_y = i(\be_{1}+\be_{-1})/\sqrt{2}$. Note that $\be_{-q} = (-1)^q (\be_q)^\ast$ for $q\in\{-1,0,1\}$. A vector $\mathbf{a}= a_x \be_x + a_y \be_y + a_z \be_z$ can thus written as $\mathbf{a}= \sum_q a_q \be_q^\ast = \sum_q (-1)^q a_q \be_{-q}$ where $a_q = \be_q \cdot \mathbf{a}$ such that $a_{\pm 1} := \tfrac{1}{\sqrt{2}} ( \mp a_x - i a_y)$ and $a_0 = a_z$ just as we had for the basis vectors above.

The dot product of two vectors $\mathbf{a}$, $\mathbf{b}$ is then given by
\begin{equation}
	\mathbf{a}\cdot\mathbf{b} = \sum_q (-1)^q a_q \be_{-q} \mathbf{b} = \sum_q (-1)^q a_q b_{-q} = \sum_q a_q b_q^\ast \,.
\end{equation}
For the cross product we note that the usual rule $\mathbf{a}\times(\mathbf{b} \times \mathbf{c}) = (\mathbf{a} \cdot \mathbf{c}) \mathbf{b} - (\mathbf{a} \cdot \mathbf{b}) \mathbf{c}$ still holds.

Just as any other vectorial quantity, the dipole operator can then be written as $\vct{d} = \sum_q d_q \be_q^\ast$ where $d_q = d_q^{(+)} + d_q^{(-)}$. Using the Wigner-Eckart Theorem~\cite{steckquantumoptics,cohenQM} we can write
\begin{subequations}
\begin{align}
	d_q^{(+)} 
		&= \sum_{m_g,m_e} \bkew{J_g\: m_g}{d_q}{J_e\: m_e} \ketbra{J_g\: m_g}{J_e\: m_e} 
		= \bkewreduced{J_g}{\vct{d}}{J_e} \sum_{m_g,m_e} \braket{J_g\: m_g}{J_e\: m_e; 1\: q} \ketbra{J_g\: m_g}{J_e\: m_e} \,,
\intertext{and, using $d_q^{(-)} = (-1)^q (d_q^{(+)})^\dagger$,}
	d_q^{(-)}& = \bkewreduced{J_g}{\vct{d}}{J_e} \sum_{m_g,m_e} (-1)^q \braket{J_g\: m_g}{J_e\: m_e; 1\: -q} \ketbra{J_e\: m_e}{J_g\: m_g} \,.
\end{align}
\end{subequations}
Here $\braket{J_g\: m_g}{J_e\: m_e; 1\: \pm q}=\braket{J_e\: m_e; 1\: \pm q}{J_g\: m_g}$ are the Clebsch-Gordan coefficients for a fine-structure transition between states $\ket{J_e\: m_e}$ and $\ket{J_g\: m_g}$ while $\bkewreduced{J_g}{\vct{d}}{J_e}$ is the (real) reduced matrix element for the whole $J_g \leftrightarrow J_e$ transition and will be included in the coupling term~$\Omega_L$. 

For the $J_g=1/2 \leftrightarrow J_e=1/2$ configuration discussed in Sect.~\ref{sec:_4lvl_examples} we set $\ket{J_g=1/2, m_g=\pm1/2} =: \ket{g_\pm}$ and two excited states $\ket{J_e=1/2, m_e=\pm1/2} =: \ket{e_\pm}$. The corresponding Clebsch-Gordan coefficients are displayed in Fig.~\ref{fig:_4lvlsetup} such that
\begin{subequations}\label{eq:_dipoleOP_4lvl_0comp}
\begin{align}
	d_{\pm1}^{(-)} &= \mp \sqrt{\tfrac{2}{3}} \bkewreduced{J_g}{\vct{d}}{J_e}  \mathcal{S}_{e_\pm g_\mp}
	\,,\\
	d_{0}^{(-)} &= \sqrt{\tfrac{1}{3}} \bkewreduced{J_g}{\vct{d}}{J_e} \left( \mathcal{S}_{e_+ g_+} - \mathcal{S}_{e_- g_-} \right)
	\,.
\end{align}	
\end{subequations}

Just as we can write $\vct{d}^{(\pm)}:=\sum_q d_q^{(\pm)} \be_q^\ast$ (such that $(\vct{d}^{(-)})^\dagger = \vct{d}^{(+)}$) we can also write the electric field as $\vct{E}=\vct{E}^{(+)} + \vct{E}^{(-)}$ where $\vct{E}^{(\pm)}=\sum_q E_q^{(\pm)} \be_q^\ast$ and
\begin{subequations}
\begin{align}
	E_q^{(+)} &= \tfrac{1}{2} \mathcal{E} \epsilon_q e^{i (\bk \cdot \vct{R} - \omega t)} \,,
	\\
	E_q^{(-)} &= (-1)^q (\vct{E}_{-q}^{(+)})^\dagger 
			= \tfrac{1}{2} \mathcal{E}^\ast \epsilon_{q} e^{-i (\bk \cdot \vct{R} - \omega t)}	\,.
\end{align}	
\end{subequations}
Here $\mathcal{E}$ is the amplitude of the field while $\epsilon_q$ describes the relative component along the $q$-direction. If the field is propagating along the quantisation axis, \ie $\bk = \omega_L \bkappa/c$ with $\bkappa \parallel \be_z=\be_0$, we get $E_0^{(\pm)}=0$. For the magnetic field we use $\vct{B}^{(\pm)} = \bkappa \times \vct{E}^{(\pm)}/c$.

In the rotating wave approximation we get the coupling term $-\vct{d}\cdot\vct{E} = - \vct{d}^{(-)} \vct{E}^{(+)} - \vct{d}^{(+)}\cdot \vct{E}^{(-)}$ with
	\begin{equation}
		-\vct{d}^{(-)} \vct{E}^{(+)} =
		\tfrac{1}{2} \hbar \Omega_L e^{i \bk \cdot \vct{R}} 
		\sum_q \epsilon_{-q} \sum_{m_g,m_e} \braket{J_g\: m_g}{J_e\: m_e; 1\: -q} \ketbra{J_e\: m_e}{J_g\: m_g} \,,
	\end{equation}
where we set $\hbar \Omega_L := - \bkewreduced{J_g}{\vct{d}}{J_e} \mathcal{E}$. The sum over $m_e=-J_e, -J_e+1, \dots, J_e$ can be evaluated easily as the Clebsch-Gordan coefficient $\braket{J_g\: m_g}{J_e\: m_e; 1\: -q}$ is non-zero only if $m_e=m_g+q$. This is why a laser-beam polarised as $\mathbf{E}^{(+)} = E_{-1}^{(+)} \be_{-1}^\ast$ induces $\sigma_+$ transitions $\ket{m_g}\rightarrow \ket{m_e=m_g+1}$ and vice versa for $\mathbf{E}^{(+)} = E_{1}^{(+)} \be_{1}^\ast$, \cf Eq.~\eqref{eq:_Fields_Example4}.

For a beam propagating along the $z$-direction we use the atom-laser Hamiltonian given in Eq.~\eqref{eq:_Hatomlaser} and set $\bkappa = \vartheta \be_z$ with $\vartheta=+1$ ($\vartheta=-1$) for propagation in the positive (negative) $z$-direction to get
\begin{equation}\label{eq:_Hatomlaser_appendix}
	H_\text{AL} =
		 \Big(1 - \tfrac{\vartheta}{M c} P_0 + \tfrac{\hbar \omega}{2 M c^2} \Big) \Big( d_1^{(-)} E_{-1}^{(+)} + d_{-1}^{(-)} E_{1}^{(+)} \Big)
			+ \tfrac{\vartheta}{M c} d_0^{(-)} \Big(P_1 E_{-1}^{(+)} + P_{-1} E_{1}^{(+)}\Big)
			+ \Hc \,,
\end{equation}
with $P_0 \rightarrow p_z$ and $P_{\pm1}\rightarrow \mp \tfrac{1}{\sqrt{2}} ( p_x \pm i p_y) $ in the semi-classical limit. Using this and the dipole operators defined in Eq.~\eqref{eq:_dipoleOP_4lvl_0comp} we get the Hamiltonian used in equs.~\eqref{eq:_Hatomlaser_4lvlatom}, \eqref{eq:_Hexample2_Appendix} and~\eqref{eq:_Hexample3_Appendix}.

\section{Evolution of atomic states for examples given in Sect.~\ref{sec:_4lvl_examples}}\label{appendix:_OptBloch}

In Sect.~\ref{sec:_4lvl_examples} we introduce a four-level system which is coupled to a magnetic field as well as one or two laser beams. To describe the evolution of the internal states we use an approach similar to the optical Bloch-equations used for a two-level system. Note, however, that the interpretation of the atom as a spin-1/2 system is no longer valid in this case.

\subsection{Hamiltonian and evolution for the example in Sect.~\ref{sec:_Ex_sigmaP_and_Bfield}}\label{appendix:_OptBloch_Ex1}

The Hamiltonian describing the interaction between the atom, the laser field coupling the states~$\ket{g_-}$ and~$\ket{e_+}$ and the magnetic field coupling~$\ket{g_+}$ and~$\ket{g_-}$ has been given in Eq.~\eqref{eq:_Hatomlaser_4lvlatom}. To evaluate the evolution of the internal states we use the Hamiltonian without the kinetic and vacuum contributions,
\begin{multline}\label{eq:_Hexample2_Appendix}
	H=- \hbar \delta \big( \mathcal{S}_{e_+ e_+} + \mathcal{S}_{e_- e_-} \big) 
	- i \frac{\hbar \Omega_B}{2} \big(\mathcal{S}_{g_- g_+} - \mathcal{S}_{g_+ g_-}\big) 
	\\
	+ \frac{\Omega_+ \hbar}{2}\Big( \mathcal{S}_{e_+ g_-} e^{i k z} + \Hc\Big)
	+  \frac{\hbar}{2}\Big( \Omega_0 \left( \mathcal{S}_{e_+ g_+} - \mathcal{S}_{e_- g_-} \right) e^{i k z} + \Hc\Big)
	\,,
\end{multline}
where we defined $\Omega_+(\zeta) := \Omega_L(\zeta) (1 - p_z/(Mc) + \hbar\omega_L/(2Mc^2))$, and $\Omega_0(\zeta) := \Omega_L(\zeta) (p_x + i p_y)/(2Mc)$. Note that $\Omega_0$ is complex. The evolution equation for the operators $\mathcal{S}_{e_+ g_-} = \ketbra{e_+}{g_-}$ \etc are then given by $\tder{t}{} \mathcal{S}_{e_+ g_-} = \tfrac{i}{\hbar} [H, \mathcal{S}_{e_+ g_-}]$.

The contribution from the vacuum Hamiltonian is included through the spontaneous decay rates. The coefficients in Fig.~\ref{fig:_4lvlsetup} also show how the total spontaneous decay from the excited to the ground state manifold branches, \ie $\Gamma_{e_+\rightarrow g_+} = \Gamma_{e_-\rightarrow g_-} = \Gamma/3$, $\Gamma_{e_+\rightarrow g_-} = \Gamma_{e_-\rightarrow g_+} = 2 \Gamma/3$. This way we find~\cite{kaiser1991mechanical}
\begin{subequations}\label{eq:_DecayRates-4lvl}
\begin{align}
	\big(\tder{t}{} {\mathcal{S}}_{e_+ e_+}\big)_\text{sp} &= 
	- \Gamma \mathcal{S}_{e_+ e_+}	
	\,,\\
	\big(\tder{t}{} {\mathcal{S}}_{e_- e_-}\big)_\text{sp} &= - \Gamma \mathcal{S}_{e_- e_-} 
	\,,\\
	\big(\tder{t}{} {\mathcal{S}}_{g_+ g_+}\big)_\text{sp} &= 
	\tfrac{1}{3} \Gamma \big(\mathcal{S}_{e_+ e_+} + 2 \mathcal{S}_{e_- e_-}\big)
	\,,\\
	\big(\tder{t}{} {\mathcal{S}}_{g_- g_-}\big)_\text{sp} &= \tfrac{1}{3} \Gamma \big(\mathcal{S}_{e_- e_-} + 2 \mathcal{S}_{e_+ e_+}\big)
	\,,\\
	\big(\tder{t}{} {\mathcal{S}}_{e_+ g_\pm}\big)_\text{sp} &= 
	- \tfrac{1}{2} \Gamma \mathcal{S}_{e_+ g_\pm}
	\,,\\
	\big(\tder{t}{} {\mathcal{S}}_{e_- g_\pm}\big)_\text{sp} &= - \tfrac{1}{2} \Gamma \mathcal{S}_{e_- g_\pm}
	\,,\\
	\big(\tder{t}{} {\mathcal{S}}_{e_+ e_-}\big)_\text{sp} &= -\Gamma \mathcal{S}_{e_+ e_-}
\end{align}
\end{subequations}

Using $\ew{\tder{t}{} S_{g_- e_+}} = \ew{\tder{t}{} S_{e_+ g_-}}^\ast$ and $\ew{S_{e_+ e_+}} +\ew{S_{e_- e_-}} + \ew{S_{g_+ g_+}} + \ew{S_{g_- g_-}} = 1$ we still need to solve 15~coupled equations to describe the average internal dynamics of the 4-level system. Defining
\begin{subequations}\label{eq:_Def-uvw_4lvl_ex1}\begin{align}
	u_{e_i g_j} + i v_{e_i g_j} &:= \ew{S_{e_i g_j}} e^{ik z}
	\,, \\
	u_{e_+ e_-} + i v_{e_+ e_-} &:= \ew{S_{e_+ e_-}}
	\,, \\
	u_{g_+ g_-} + i v_{g_+ g_-} &:= \ew{S_{g_+ g_-}}
	\,, \\
	w_{e_+ g_-} &:= \ew{S_{e_+ e_+}} - \ew{S_{g_- g_-}} 
	\,, \\
	w_{e_- g_+} &:= \ew{S_{e_- e_-}} - \ew{S_{g_+ g_+}} 
	\,, \\
	w_{g_+ g_-} &:= \ew{S_{g_+ g_+}} - \ew{S_{g_- g_-}} 
	\,,
\end{align}\end{subequations}
for $i,j \in \{+,-\}$ and setting $\Delta := \omega_L-\omega_A- k \dot{z}$ these evolution equations read
\begin{align*}
	\dot{w}_{e_+ g_-} &=  
		2 \Omega_+ v_{e_+ g_-} + \Omega_B u_{g_+ g_-} + \re\Omega_0 \big(v_{e_+ g_+} - v_{e_- g_-}\big) + \im\Omega_0 \big(u_{e_+ g_+} - u_{e_- g_-}\big) 
	\notag \\ &\qquad 
		+ \tfrac{1}{6}\Gamma \big( w_{e_- g_+} - 7 w_{e_+ g_-} + 4 w_{g_+ g_-}\big) - \Gamma/2 
\,,\\
	\dot{w}_{e_- g_+} &= 
		- \Omega_B u_{g_+ g_-} + \re\Omega_0 \big(v_{e_+ g_+} - v_{e_- g_-}\big) + \im\Omega_0 \big(u_{e_+ g_+} - u_{e_- g_-}\big) 
	\notag \\ &\qquad 
		+ \tfrac{1}{6} \Gamma \big(w_{e_+ g_-} - 7 w_{e_- g_+} - 4 w_{g_+ g_-}\big) - \Gamma/2 
\,,\\
	\dot{w}_{g_+ g_-} &= 
		\Omega_+ v_{e_+ g_-} + 2 \Omega_B u_{g_+ g_-} - \re\Omega_0 \big(v_{e_- g_-} + v_{e_+ g_+}\big) - \im\Omega_0 \big( u_{e_- g_-} + u_{e_+ g_+}\big) 
	\notag \\ &\qquad 
		+ \tfrac{1}{3} \Gamma \big(w_{e_- g_+} - w_{e_+ g_-} + w_{g_+ g_-}\big)
\,,\\ 
\notag \\
	\dot{u}_{e_+ g_-} &= \Delta  v_{e_+ g_-} - \tfrac{1}{2}\Omega_B u_{e_+ g_+} - \tfrac{1}{2}\re\Omega_0 \big(v_{e_+ e_-} - v_{g_+ g_-}\big) + \tfrac{1}{2}\im\Omega_0 \big(u_{e_+ e_-} + u_{g_+ g_-}\big) - \tfrac{1}{2}\Gamma u_{e_+ g_-} 
\,,\\
	\dot{u}_{e_- g_+} &=  \Delta  v_{e_- g_+} + \tfrac{1}{2} \Omega_B u_{e_- g_-} - \tfrac{1}{2}\re\Omega_0 \big(v_{e_+ e_-} + v_{g_+ g_-}\big) - \tfrac{1}{2}\im\Omega_0 \big(u_{e_+ e_-} + u_{g_+ g_-}\big) - \tfrac{1}{2}\Gamma  u_{e_- g_+} 
\,,\\
	\dot{v}_{e_+ g_-} &= - \Delta  u_{e_+ g_-} - \tfrac{1}{2}\Omega_+ w_{e_+ g_-} - \tfrac{1}{2}\Omega_B v_{e_+ g_+} + \tfrac{1}{2}\re\Omega_0 \big(u_{e_+ e_-} + u_{g_+ g_-}\big) + \tfrac{1}{2}\im\Omega_0 \big(v_{e_+ e_-} + v_{g_+ g_-}\big) - \tfrac{1}{2}\Gamma  v_{e_+ g_-}  
\,,\\
	\dot{v}_{e_- g_+} &= - \Delta  u_{e_- g_+} + \tfrac{1}{2}\Omega_B v_{e_- g_-} - \tfrac{1}{2} \re\Omega_0 \big(u_{e_+ e_-} + u_{g_+ g_-}\big) + \tfrac{1}{2}\im\Omega_0 \big(v_{e_+ e_-} + v_{g_+ g_-}\big) - \tfrac{1}{2} \Gamma  v_{e_- g_+}
\,,\\
\notag \\
	\dot{u}_{e_+ g_+} &=  \Delta  v_{e_+ g_+} + \tfrac{1}{2} \Omega_+ v_{g_+ g_-} + \tfrac{1}{2} \Omega_B u_{e_+ g_-} + \tfrac{1}{2}\im\Omega_0 \big( w_{g_+ g_-} - w_{e_+ g_-} \big) - \tfrac{1}{2}\Gamma u_{e_+ g_+}
\,,\\
	\dot{u}_{e_- g_-} &= \Delta v_{e_- g_-} - \tfrac{1}{2} \Omega_+ v_{e_+ e_-} - \tfrac{1}{2}\Omega_B u_{e_- g_+} + \tfrac{1}{2}\im\Omega_0 \big(w_{e_- g_+} + w_{g_+ g_-}\big) - \tfrac{1}{2}\Gamma u_{e_- g_-} 
\,,\\
	\dot{v}_{e_+ g_+} &= - \Delta u_{e_+ g_+} + \tfrac{1}{2}\Omega_+ u_{g_+ g_-} + \tfrac{1}{2}\Omega_B v_{e_+ g_-} + \tfrac{1}{2}\re\Omega_0 \big(w_{g_+ g_-} - w_{e_+ g_-}\big) - \tfrac{1}{2} \Gamma v_{e_+ g_+}
\,,\\
	\dot{v}_{e_- g_-} &= - \Delta u_{e_- g_-} - \tfrac{1}{2}\Omega_+ u_{e_+ e_-} - \tfrac{1}{2} \Omega_B v_{e_- g_+} + \tfrac{1}{2}\re\Omega_0 \big(w_{e_- g_+} + w_{g_+ g_-}\big) - \tfrac{1}{2} \Gamma v_{e_- g_-}
\,,\\
\notag \\
	\dot{u}_{e_+ e_-} &= \tfrac{1}{2}\Omega_+ v_{e_- g_-} + \tfrac{1}{2}\re\Omega_0 \big(v_{e_- g_+} - v_{e_+ g_-}\big) + \tfrac{1}{2}\im\Omega_0 \big( u_{e_- g_+} - u_{e_+ g_-}\big) - \Gamma  u_{e_+ e_-} 
\,,\\
	\dot{u}_{g_+ g_-} &= -  \tfrac{1}{2} \Omega_+ v_{e_+ g_+} -  \tfrac{1}{2} \Omega_B w_{g_+ g_-} + \tfrac{1}{2}\re\Omega_0 \big(v_{e_- g_+} - v_{e_+ g_-}\big) + \tfrac{1}{2}\im\Omega_0 \big(u_{e_- g_+} - u_{e_+ g_-}\big)
\,,\\
	\dot{v}_{e_+ e_-} &= \tfrac{1}{2} \Omega_+ u_{e_- g_-} + \tfrac{1}{2}\re\Omega_0 \big(u_{e_- g_+} + u_{e_+ g_-}\big) - \tfrac{1}{2}\im\Omega_0 \big(v_{e_- g_+} + v_{e_+ g_-}\big) - \Gamma  v_{e_+ e_-}
\,,\\
	\dot{v}_{g_+ g_-} &= - \tfrac{1}{2} \Omega_+ u_{e_+ g_+} + \tfrac{1}{2} \re\Omega_0 \big(u_{e_- g_+} + u_{e_+ g_-}\big) - \tfrac{1}{2}\im\Omega_0 \big(v_{e_- g_+} + v_{e_+ g_-}\big)
\,.
\end{align*}	

In the examples presented in Figs.~\ref{fig:_fourlvl_singleLaserB_SS_uv} to~\ref{fig:_fourlvl_singleLaserB_Ex2} we assumed the atom to be momentarily at rest such that $\Omega_0=0$ and $\tder{t}{}=\tpder{t}{}$. Even for moving atom the terms~$\sim\Omega_0$ are suppressed as they are proportional to~$v_{x,y}/c$.

\subsection{Hamiltonian and evolution for the example in Sect.~\ref{sec:_Ex_sigmaPM_and_Bfield}}\label{appendix:_OptBloch_Ex2}
The example described in Sect.~\ref{sec:_Ex_sigmaPM_and_Bfield} includes two counter-propagating laser beams, each driving a different transition in the four-level setup shown in Fig.~\ref{fig:_4lvlsetup}. The intensity of both beams and the magnetic coupling~$\Omega_B$ are set constant, the time-variation is introduced by a difference in the relative frequencies between the beams.

Using the fields given in~\eqref{eq:_Fields_Example4} and the atom-light Hamiltonian from~\eqref{eq:_Hatomlaser_appendix} we obtain the Hamiltonian governing the internal atomic dynamics,
\begin{multline}\label{eq:_Hexample3_Appendix}
	H= \tfrac{\mathbf{p}^2}{2M} - \hbar \delta \big( \mathcal{S}_{e_+ e_+} + \mathcal{S}_{e_- e_-} \big) - i \frac{\hbar}{2} \Omega_B \big(\mathcal{S}_{g_- g_+} - \mathcal{S}_{g_+ g_-}\big)
 	\\
 	+ \alpha \tfrac{\hbar \Omega_L}{2} \left( 1 - \tfrac{p_z}{Mc} + \tfrac{\hbar (\omega_L+\sigma)}{2 Mc^2}\right) \left( \mathcal{S}_{e_+ g_-} e^{i( kz - \zeta)} + \Hc \right)
 	- \tfrac{\hbar \Omega_L}{2} \left( 1 + \tfrac{p_z}{Mc} + \tfrac{\hbar \omega_L}{2 Mc^2} \right) \left( \mathcal{S}_{e_- g_+} e^{-i kz} + \Hc \right)
\\
 	+ \tfrac{\hbar \Omega_L}{4 Mc} \Big( \big( \mathcal{S}_{e_+ g_+} - \mathcal{S}_{e_- g_-} \big)
 		\Big( p_x \left( \alpha e^{i(kz-\zeta)} + e^{-ikz}\right) 
 		+ i p_y \left( \alpha e^{i(kz-\zeta)} - e^{-ikz}\right) \Big) + \Hc \Big)
 	\,,
\end{multline}
where $\zeta=\sigma t - \sigma z/c$. The vacuum contributions leading to spontaneous decay are included by the rules given in Eq.~\eqref{eq:_DecayRates-4lvl}. The real, averaged quantities $u, v, w$ are defined as given in~\eqref{eq:_Def-uvw_4lvl_ex1} with the exceptions
\begin{subequations}\label{eq:_Def-uvw_4lvl_ex2}\begin{align}
	u_{e_+ g_-} + i v_{e_+ g_-} &:= \ew{S_{e_+ g_-}} e^{i( k z - \zeta)}
	\,, \\
	u_{e_- g_+} + i v_{e_- g_+} &:= \ew{S_{e_- g_+}} e^{-i k z}
	\,, \\
	u_{e_+ g_+} + i v_{e_+ g_+} &:= \ew{S_{e_+ g_+}}
	\,, \\
	u_{e_- g_-} + i v_{e_- g_-} &:= \ew{S_{e_- g_-}}
	\,.
\end{align}\end{subequations}
Dropping terms proportional to $p_{x,y,z}/Mc$ or $\dot{z}/c$ we find that the states evolve as
\begin{align*}
	\dot{w}_{e_+ g_-} &= 
		2 \alpha  \Omega_L v_{e_+ g_-} + \Omega_B u_{g_+ g_-} + \tfrac{1}{6} \Gamma \big(w_{e_- g_+}-7 w_{e_+ g_-}+4 w_{g_+ g_-}\big) - \Gamma/2
\,, \\
	\dot{w}_{e_- g_+} &=
		-2 \Omega_L v_{e_- g_+} -u_{g_+ g_-} \Omega_B + \tfrac{1}{6} \Gamma \big(-7 w_{e_- g_+}+w_{e_+ g_-}-4 w_{g_+ g_-}\big) -\Gamma/2
\,, \\
	\dot{w}_{g_+ g_-} &=
		\Omega_L \big(v_{e_- g_+}+\alpha v_{e_+ g_-}\big) + 2 \Omega_B u_{g_+ g_-}+ \tfrac{1}{3} \Gamma \big(w_{e_- g_+}-w_{e_+ g_-}+w_{g_+ g_-}\big)
\,, \\
\notag \\
	\dot{u}_{e_+ g_-} &=
		(\delta+\sigma) v_{e_+ g_-}+\tfrac{1}{2} \Omega_B \big(\sin(k z-\zeta) v_{e_+ g_+}-\cos(k z-\zeta) u_{e_+ g_+}\big) -\tfrac{1}{2}\Gamma u_{e_+ g_-}
\,, \\
	\dot{u}_{e_- g_+} &= 
		\delta v_{e_- g_+} +\tfrac{1}{2} \Omega_B \big(\cos(k z) u_{e_- g_-}+\sin(k z) v_{e_- g_-}\big) - \tfrac{1}{2} \Gamma u_{e_- g_+}		
\,, \\
	\dot{v}_{e_+ g_-} &= 
		-(\delta+\sigma) u_{e_+ g_-} -\tfrac{1}{2} \alpha \Omega_L  w_{e_+ g_-} -\tfrac{1}{2}\Omega_B \big(\sin(k z-\zeta) u_{e_+ g_+}+\cos(k z-\zeta) v_{e_+ g_+}\big) -\tfrac{1}{2}\Gamma v_{e_+ g_-}
\,, \\
	\dot{v}_{e_- g_+} &= 
		-\delta u_{e_- g_+} + \tfrac{1}{2} \Omega_L w_{e_- g_+} + \tfrac{1}{2} \Omega_B \big(\cos(k z) v_{e_- g_-}-\sin(k z) u_{e_- g_-}\big) - \tfrac{1}{2}\Gamma v_{e_- g_+}
\,, \\
\notag\\
	\dot{u}_{e_+ g_+} &= 
		\delta v_{e_+ g_+} + \tfrac{1}{2} \Omega_L \big(-\sin(k z) u_{e_+ e_-}+\alpha \sin(k z-\zeta) u_{g_+ g_-}-\cos(k z) v_{e_+ e_-}+\alpha \cos(k z-\zeta) v_{g_+ g_-}\big) 
	\notag \\ &\qquad 
		+ \tfrac{1}{2} \Omega_B \big(\cos(k z-\zeta) u_{e_+ g_-}+\sin(k z-\zeta) v_{e_+ g_-}\big) - \tfrac{1}{2}\Gamma u_{e_+ g_+}
\,, \\
	\dot{u}_{e_- g_-} &= 
		\delta v_{e_- g_-} + \tfrac{1}{2} \Omega_L \big(\sin(k z) u_{g_+ g_-}- \alpha \sin(k z-\zeta) u_{e_+ e_-} - \alpha \cos(k z-\zeta) v_{e_+ 	e_-}+\cos(k z) v_{g_+ g_-}\big)
	\notag \\ & \qquad
		+ \tfrac{1}{2} \Omega_B \big(\sin(k z) v_{e_- g_+}-\cos(k z) u_{e_- g_+}\big) -\tfrac{1}{2} \Gamma u_{e_- g_-}		
\,, \\
	\dot{v}_{e_+ g_+} &= 
		-\delta u_{e_+ g_+} + \tfrac{1}{2} \Omega_L \big(\cos(k z) u_{e_+ e_-}+\alpha \cos(k z-\zeta) u_{g_+ g_-}-\sin(k z) v_{e_+ e_-}-\alpha \sin(k z-\zeta) v_{g_+ g_-}\big) 
	\notag \\ &\qquad
		+ \tfrac{1}{2} \Omega_B \big(\cos(k z-\zeta) v_{e_+ g_-}-\sin(k z-\zeta) u_{e_+ g_-}\big) - \tfrac{1}{2}\Gamma v_{e_+ g_+}
\,, \\
	\dot{v}_{e_- g_-} &= 
		-\delta u_{e_- g_-} + \tfrac{1}{2} \Omega_L \big(-\alpha \cos(k z-\zeta) u_{e_+ e_-}-\cos(k z) u_{g_+ g_-}+\alpha \sin(k z-\zeta) v_{e_+ e_-}+\sin(k z) v_{g_+ g_-}\big)
	\notag \\ & \qquad
		- \tfrac{1}{2} \Omega_B \big(\sin(k z) u_{e_- g_+}+\cos(k z) v_{e_- g_+}\big) - \tfrac{1}{2} \Gamma v_{e_- g_-}
\,, \\
\notag \\ 
	\dot{u}_{e_+ e_-} &= 
		\tfrac{1}{2} \Omega_L \big(\alpha \sin(k z-\zeta) u_{e_- g_-}+\sin(k z) u_{e_+ g_+}+\alpha \cos(k z-\zeta) v_{e_- g_-}-\cos(k z) v_{e_+ g_+}\big) -\Gamma u_{e_+ e_-}	
\,, \\
	\dot{u}_{g_+ g_-} &= 
		\tfrac{1}{2} \Omega_L \big(-\sin(k z) u_{e_- g_-}+\cos(k z) v_{e_- g_-}-\alpha \sin(k z-\zeta) u_{e_+ g_+} -\alpha \cos(k z-\zeta) v_{e_+ g_+}\big) -\tfrac{1}{2} \Omega_B w_{g_+ g_-}
\,, \\
	\dot{v}_{e_+ e_-} &= 
		\tfrac{1}{2} \Omega_L \big(\alpha \cos(k z-\zeta) u_{e_- g_-}+\cos(k z) u_{e_+ g_+}-\alpha \sin(k z-\zeta) v_{e_- g_-}+\sin(k z) v_{e_+ g_+}\big) - \Gamma v_{e_+ e_-}
\,, \\
	\dot{v}_{g_+ g_-} &= 
		\tfrac{1}{2} \Omega_L \big(-\cos(k z) u_{e_- g_-}-\alpha \cos(k z-\zeta) u_{e_+ g_+}-\sin(k z) v_{e_- g_-}+\alpha \sin(k z-\zeta) v_{e_+ g_+}\big)
\,.	
\end{align*}

Note that these evolution equations are explicitly time-dependent as $\zeta = \sigma(t-z/c)$. The steady-state solutions shown in Figs.~\ref{fig:4lvl_sigmaPM_Ex2}(a) and~(b) are thus calculated with $\sigma=0$. The full solutions are calculated numerically assuming periodic boundary conditions.

%
%

%
\end{document}